\documentclass[twocolumn,notitlepage,superscriptaddress]{revtex4-1}  

\usepackage[colorlinks=true,urlcolor=blue,linkcolor=blue,citecolor=blue]{hyperref}
\usepackage{latexsym}
\usepackage{graphicx}
\usepackage{subfigure} 
\usepackage{placeins}
\usepackage{amsmath}
\usepackage{color}
\usepackage{amsfonts}
\usepackage{mathrsfs}
\usepackage{simpler-wick}

\usepackage{accents}
\makeatletter
\def\wideubar{\underaccent{{\cc@style\underline{\mskip10mu}}}}
\def\Wideubar{\underaccent{{\cc@style\underline{\mskip8mu}}}}
\makeatother
\makeatletter
\def\widebar{\accentset{{\cc@style\underline{\mskip10mu}}}}
\def\Widebar{\accentset{{\cc@style\underline{\mskip8mu}}}}
\makeatother

\begin{document}

\title{ Coincidence double-tip scanning tunneling spectroscopy }

\author{Yuehua Su}
\email{suyh@ytu.edu.cn}
\affiliation{ Department of Physics, Yantai University, Yantai 264005, People's Republic of China }

\author{Guoya Zhang}
\affiliation{ Department of Physics, Yantai University, Yantai 264005, People's Republic of China }

\author{Dezhong Cao}
\affiliation{ Department of Physics, Yantai University, Yantai 264005, People's Republic of China }

\author{Chao Zhang}
\affiliation{ Department of Physics, Yantai University, Yantai 264005, People's Republic of China }

\begin{abstract}

The development of new experimental techniques for direct measurement of many-body correlations is crucial for unraveling the mysteries of strongly correlated electron systems. In this work, we propose a coincidence double-tip scanning tunneling spectroscopy (STS) that enables direct probing of spatially resolved dynamical two-body correlations of sample electrons. Unlike conventional single-tip scanning tunneling microscopy, the double-tip STS employs a double-tip scanning tunneling microscope (STM) equipped with two independently controlled tips, each biased at distinct voltages ($V_1$ and $V_2$). By simultaneously measuring the quantum tunneling currents $I_1(t)$ and $I_2(t)$ at locations $j_1$ and $j_2$, we obtain a coincidence tunneling current correlation $\overline{\langle I_1(t) I_2(t)\rangle}$. Differentiating this coincidence tunneling current correlation with respect to the two bias voltages yields a coincidence dynamical conductance. Through the development of a nonequilibrium theory, we demonstrate that this coincidence dynamical conductance is proportional to a contour-ordered second-order current correlation function. For the sample electrons in a nearly free Fermi liquid state, the coincidence dynamical conductance captures two correlated dynamical electron propagation processes: (i) from $j_1$ to $j_2$ (or vice versa) driven by $V_1$, and (ii) from $j_2$ to $j_1$ (or vice versa) driven by $V_2$. For the sample electrons in a superconducting state, additional propagation channels emerge from the superconducting condensate, coexisting with the above normal electron propagation processes. Thus, the coincidence double-tip STS provides direct access to spatially resolved dynamical two-body correlations, offering a powerful tool for investigating strongly correlated electron systems.  

\end{abstract}


\maketitle

\section{Introduction} \label{sec1}

There are many unsolved mysteries in the field of strongly correlated electron systems, including unconventional superconductivity, itinerant magnetism, quantum spin liquids, quantum phase transitions and quantum criticality \citep{StewartNFLRMP2001, PALeeRMP2006, StewartFeSCRMP2011, ZhouYiRMP2017}. A central challenge in this field is the development of innovative experimental methods and advanced techniques to uncover the fundamental physics governing these enigmatic phenomena. Recently, several coincidence detection techniques have been proposed as promising tools for direct measurement of two-body correlations of strongly correlated electron systems \citep{SuZhang2020, SucINS2021, DevereauxPRB2023}. For instance, the coincidence angle-resolved photoemission spectroscopy (cARPES) can enable direct measurement of two-body correlations in particle-particle channel, offering a pathway to investigate the mechanism of unconventional superconductivity \citep{SuZhang2020, DevereauxPRB2023}. Similarly, the newly proposed coincidence inelastic neutron scattering (cINS) provides a technique to detect two-spin correlations, opening new avenues for exploring quantum spin liquid physics \citep{SucINS2021}.    

\begin{figure}[th]
\includegraphics[width=0.9\columnwidth]{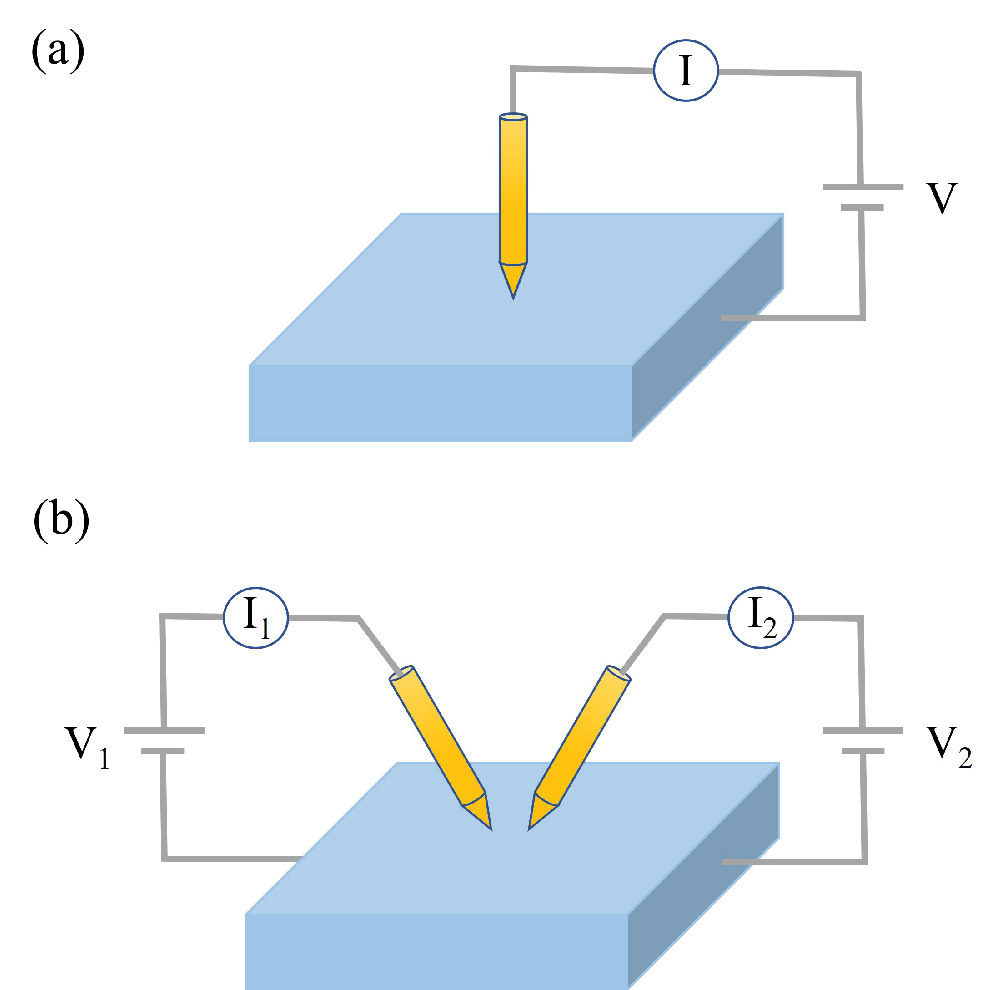}  
\caption{ Schematic diagrams of (a) a single-tip STM with an ammeter and an applied bias voltage $V$, and (b) a double-tip STM with two ammeters and two bias voltages $V_1$ and $V_2$. }
\label{fig1}
\end{figure}

In this work, we propose a new coincidence technique for spatially resolved measurement of two-body correlations,  a coincidence double-tip scanning tunneling spectroscopy (STS). This technique utilizes a double-tip scanning tunneling microscope (STM)  setup, constituting a fundamental extension of conventional single-tip STM beyond single-particle measurements. Schematic illustrations of both single-tip and double-tip STM configurations are provided in Fig.~\ref{fig1}. Since its invention, the single-tip STM has played a pivotal role in advancing atomic-resolution topography and local electronic structure studies across diverse materials in various fields of science and technology \citep{BinningHeinrichRMP1987, FischerRMP2007, ChenCJBook2008}. In this work, we focus on its extension in probing the local electronic structure of materials, scanning tunneling spectroscopy (STS). The single-tip STS measures the spatially resolved quantum tunneling current $I_j(V)$, where $j$ denotes the spatial location and $V$ is the applied bias voltage. By differentiating the tunneling current with respect to the bias voltage, the single-tip STS yields the local dynamical conductance   
\begin{equation}
\sigma_j(V) = \frac{d}{dV} I_j(V). \label{eqn1.1}
\end{equation}  
Introduce an effective current operator
\begin{equation}
J_A(j\sigma; t, t_1; \omega) = -i c_{j\sigma}(t) c^\dag_{j\sigma} (t_1) e^{-i\omega (t_1 - t)} + h.c. , \label{eqn1.2}
\end{equation}
where $c_{j\sigma}$ ($c^\dag_{j\sigma}$) are annihilation (creation) operators for the sample electrons at location $j$ with spin $\sigma$. In a nonequilibrium theory, $\sigma_j(V)$ can be shown to follow
\begin{equation}
\sigma_j(V) \sim i \sum_{\sigma} \int_c d t_1 \langle T_c J_A(j\sigma; t, t_1; -\frac{1}{\hbar} e V) \rangle . \label{eqn1.3}
\end{equation}
Here $T_c$ is a contour-time ordering operator defined along a time contour $t_i\rightarrow t\rightarrow t_i$, where $t_i$ is the initial time and $t$ represents the observation time.  The integral $\int_c d t_1$ is performed along this closed time contour.  In our study, we take $t_i\rightarrow -\infty$. $\langle O \rangle$ defines the ensemble statistical average of an observable operator $O$. It can be shown that $\sigma_j(V)$ is proportional to the local electronic single-particle spectral function (or density of states). Therefore, the single-tip STS can provide crucial insights into the single-particle electronic structure at specific locations on the sample surface.  

In a coincidence double-tip STS measurement, two independent tips enable simultaneous electron tunnelings between sample locations $j_1$ and $j_2$ to their respective probes. The tunneling currents measured at each probe at time $t$ are denoted as $I^{(1)}_1(t, n)$ and $I^{(1)}_2 (t, n)$, where the integer $n$ represents different measurements. We introduce a coincidence tunneling current observable $I_d^{(2)}$ in experiment, which captures the simultaneous tunneling current events by a post-experiment coincidence counting method \citep{SuCaoarXiv2024} as 
\begin{equation}
I_d^{(2)} = \frac{1}{N_n} \sum_{n} I^{(1)}_1(t, n) \cdot I^{(1)}_2(t, n) , \label{eqn1.4}
\end{equation} 
where $N_n$ is the total number of measurements. Theoretically, the coincidence tunneling current observable $I_d^{(2)}$ can be described by a statistical average correlation function $I_d^{(2)}=\overline{\langle I_1(t) I_2(t)\rangle}$, where $I_1(t)$ and $I_2(t)$ represent the local current operators at locations $j_1$ and $j_2$, respectively. We can then introduce a coincidence dynamical conductance as 
\begin{equation}
\sigma_{j_1 j_2}(V_1, V_2) = \frac{d}{d V_2} \frac{d}{d V_1} \overline{\langle I_1(t) I_2(t)\rangle} . \label{eqn1.5}
\end{equation} 
It can be shown in a nonequilibrium theory that
\begin{equation}
\sigma_{j_1 j_2}(V_1, V_2) \sim -\sum_{\sigma_1 \sigma_2} G_J^{(2)} (j_1\sigma_1, j_2\sigma_2; -\frac{1}{\hbar} e V_1,-\frac{1}{\hbar} e V_2 ) , \label{eqn1.6}
\end{equation}
where $G_J^{(2)}$ is a contour-time relevant second-order current correlation function defined as 
\begin{eqnarray}
&& G_J^{(2)} (j_1\sigma_1, j_2\sigma_2; \omega_1, \omega_2 ) \notag \\
&=& \iint_c d t_2 d t_1 \langle T_c J_{1A}(j_1\sigma_1; t, t_1; \omega_1) J_{2A}(j_2\sigma_2; t, t_2; \omega_2)  \rangle . \,\,\quad \label{eqn1.7}
\end{eqnarray}
Here $J_{1A}$ and $J_{2A}$ are two effective current operators defined by
\begin{eqnarray}
&& J_{1A}(j_1\sigma_1; t, t_1; \omega_1) = -i c_{j_1\sigma_1}(t) c^\dag_{j_1\sigma_1} (t_1) e^{-i\omega_1 (t_1 - t)} + h.c. , \notag \\
&& J_{2A}(j_2\sigma_2; t, t_2; \omega_2) = -i c_{j_2\sigma_2}(t) c^\dag_{j_2\sigma_2} (t_2) e^{-i\omega_2 (t_2 - t)} + h.c.  \notag \\  
&&  \label{eqn1.8}
\end{eqnarray}
By comparing $\sigma_j(V)$ with $\sigma_{j_1 j_2}(V_1, V_2)$, it shows that while $\sigma_j(V)$ provides the spatially resolved single-particle dynamical information, $\sigma_{j_1 j_2}(V_1, V_2)$ reveals the spatially resolved dynamical two-body correlation functions. Therefore, the coincidence double-tip STS is a promising technique for studying the spatially resolved dynamical two-body correlations of the esoteric strongly correlated electron systems. 

Notably, multiple-tip STM techniques have been developed over nearly three decades \citep{ByersPRL741995, NiuPRB1995, SHIRAKI2001633, NakayamaAdM201200257, BertRevSI2018, LeeuwenhoekPRB2020, MaartenNature2020}. These techniques enable nanoscale charge transport measurements like a multimeter at the nanoscale \citep{SHIRAKI2001633, NakayamaAdM201200257, BertRevSI2018}, probing the single-particle charge-current relevant Green's functions of sample electrons \citep{ByersPRL741995, NiuPRB1995}. A new approach employing double-tip STM to probe disorder-induced current-current correlations has been proposed recently for investigating electron diffusion propagation in disordered metals \citep{LeeuwenhoekPRB2020}.  Since such current-current correlations originate from disorder-induced statistical effects \citep{MirlinPRE1997, MIRLIN2000259}, they can not capture the intrinsic dynamical two-body correlations of sample electrons that our coincidence double-tip STS  probes.

This article is organized as follows. In Sec. \ref{sec2}, we revisit the conventional single-tip STS. We first review the basic principle of the single-tip STS within the framework of the equilibrium linear-response theory in Sec. \ref{sec2.1}, and we then develop a new nonequilibrium theory for the single-tip STS in Sec. \ref{sec2.2}. In Secs. \ref{sec2.3} and \ref{sec2.4}, we present the local dynamical conductances for the Fermi liquid state and the mean-field superconducting state, respectively. 

In Sec. \ref{sec3}, we propose the coincidence double-tip STS. In Sec. \ref{sec3.1}, we develop a nonequilibrium theory for the coincidence double-tip STS, which is extended from the nonequilibrium theory developed in Sec. \ref{sec2.2} for the single-tip STS. The coincidence dynamical conductances for the Fermi liquid state and the mean-field superconducting state are derived in Secs. \ref{sec3.2} and \ref{sec3.3}, respectively.

We further propose an extended double-tip STS in Sec. \ref{sec4}, generalizing the previous simultaneous current measurement (Sec. \ref{sec3}) to the cases where the two tunneling currents are measured at different times. In Sec. \ref{sec4.1}, we develop the corresponding nonequilibrium theory, while in Sec. \ref{sec4.2} we illustrate its application by the coincidence dynamical conductance for the Fermi liquid state. Some additional derivation details are provided in the Appendix sections.

\section{Single-tip scanning tunneling spectroscopy} \label{sec2}

\begin{widetext}

\begin{figure}[ht]
\includegraphics[width=0.9\columnwidth]{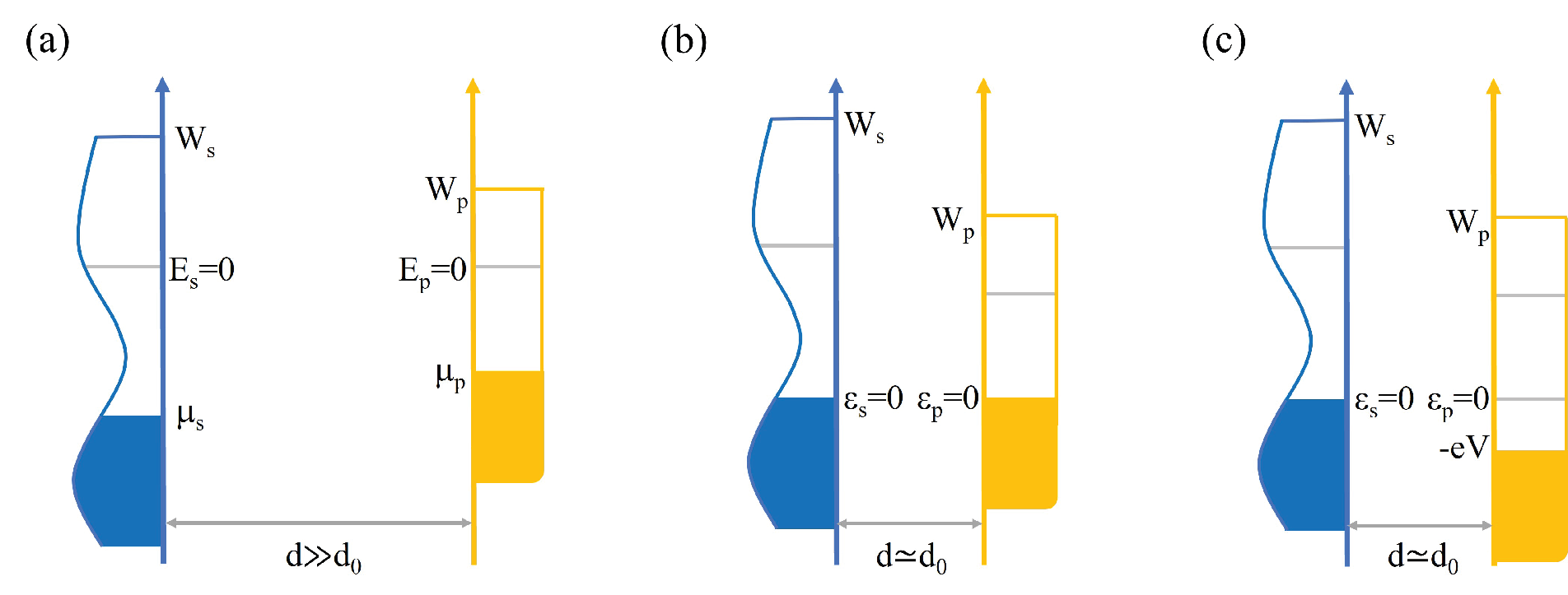}  
\caption{ Schematic illustration of quantum tunneling current formation. Panels (a)-(c) show the single-particle spectral functions of the sample (left) and tip (right), with color-shaded regions indicating occupied ones.  Here, $W_s$ and $W_p$ denote the work done by the surface electric field as an electron tunnels across the sample or tip surface. The single-electron energies of the sample and tip are given by $E_s$ and $E_p$, while $\mu_s$ and $\mu_p$ represent their respective chemical potentials. The effective single-electron energies, $\varepsilon_s = E_s - \mu_s$ and  $\varepsilon_p = E_p - \mu_p$, are defined relative to these chemical potentials. (a) At sample-tip separation $d\gg d_0$ (tunneling distance), the two systems remain decoupled  with zero net tunneling current. (b) When the sample and tip are brought to tunneling proximity ($d\simeq d_0$), their Fermi energies equilibrate, also with zero net tunneling current. (c) Application of a tip bias voltage $V$ shifts the tip single-electron energies by $-eV$, creating a nonequilibrium condition that drives finite tunneling current.}
\label{fig2}
\end{figure}
\end{widetext}

In this section, we will revisit how the single-tip STS works in study of the local electronic structure of sample material. Fig.~\ref{fig1} (a) shows a schematic diagram of the single-tip STM. When the tip approaches the sample surface within tunneling distance (typically sub-nanometer scale) and a finite bias voltage is applied, a quantum tunneling current emerges between the sample and tip. This tunneling phenomenon is further illustrated schematically in Fig.~\ref{fig2}.

Consider the configuration shown in Fig.~\ref{fig2} (a), where the sample-tip separation $d$ significantly exceeds the characteristic tunneling distance $d_0\simeq 5-10$ \AA \ \citep{FischerRMP2007}. Under this condition, the electrons of the sample and tip remain decoupled, preventing quantum tunneling across the large-spacing vacuum barrier. For the electrically neutral sample and tip, their chemical potentials $\mu_s$ and $\mu_p$ differ, with the respective work functions approximately defined by $\phi_s = W_s - \mu_s$ and $\phi_p= W_p - \mu_p$ \citep{AshcroftMermin1976}. Here $W_s$ and $W_p$ are contributions from the surface effects. The color-shaded regions in Fig.~\ref{fig2} represent occupied single-electron spectral distributions. For our subsequent study, we assume that the tip consists of a metallic material whose electrons are in a nearly free Fermi liquid state.

When the sample and tip approach within the tunneling distance range ($d\simeq d_0$),  electrons of minute quantity momentarily tunnel from the side with the higher Fermi level to the side with the lower Fermi level. This process equilibrates the Fermi energies of the sample and tip upon reaching final equilibrium state \citep{AshcroftMermin1976}. To quantitatively describe this equilibrium state, we define the effective single-electron energies $\varepsilon_s = E_s - \mu_s$ and  $\varepsilon_p = E_p - \mu_p$, where $E_s$ and $E_p$ denote the single-electron energies of the sample and tip, respectively, as illustrated in Fig.~\ref{fig2} (b).  

Application of a bias voltage $V$ modifies this equilibrium configuration, as shown in Fig.~\ref{fig2} (c). The tip single-electron energies experience a uniform shift of $-eV$, creating an energy window $(-eV, 0)$ where the sample single-electron spectral distributions remain occupied while the corresponding tip single-electron spectral distributions become unoccupied. This nonequilibrium population change causes electrons to tunnel between the sample and tip, generating a finite quantum tunneling current. Crucially, while the unbiased system (b) maintains equilibrium, the biased configuration (c) represents a nonequilibrium dynamic state. This fundamental distinction necessitates a nonequilibrium theory to appropriately describe the voltage-driven tunneling phenomena.

\subsection{Basic principle: Equilibrium linear-response theory} \label{sec2.1}

Let us review the basic principle of the single-tip STS from an equilibrium linear-response theory. Suppose at time $t<t_i$, the sample-tip system is in the unbiased equilibrium state of Fig.~\ref{fig2} (b). The sample-tip system can now be described by an ensemble density matrix 
\begin{equation}
\rho_0 = \frac{1}{Z} e^{-\beta [K_s^{(0)} + K_p^{(0)}]} , \label{eqn2.1.1}
\end{equation}
where $K_s^{(0)}$ and $K_p^{(0)}$ are defined by
\begin{eqnarray}
K_s^{(0)} &=& H_s -\mu_s \sum_{i\sigma} c^{\dag}_{i\sigma} c_{i\sigma}, \notag  \\
K_p^{(0)} &=& H_p -\mu_p \sum_{i\sigma} d^{\dag}_{i\sigma} d_{i\sigma}, \label{eqn2.1.2}
\end{eqnarray}
and the partition function $Z$ is defined as $Z=\operatorname{Tr} \{ \exp[-\beta (K_s^{(0)} + K_p^{(0)})]\}$. $H_s$ and $H_p$ are the respective Hamiltonians of the sample and tip electrons, and $c^{\dag}_{i\sigma} (c_{i\sigma})$ and $d^{\dag}_{i\sigma} (d_{i\sigma})$ are the corresponding electron creation (annihilation) operators at location $i$ with spin $\sigma$. 

At time $t_i$, a bias voltage $V$ is applied to the system, inducing a shift of $-eV$ of the tip single-electron energies. Consequently, the sample and tip Hamiltonians with respect to the respective chemical potentials transform into the following forms:  
\begin{eqnarray}
K_s^{(0)} &\rightarrow & K_s = K_s^{(0)} , \notag \\
K_p^{(0)} &\rightarrow & K_p = K_p^{(0)} - eV\sum_{i\sigma} d^{\dag}_{i\sigma} d_{i\sigma} . \label{eqn2.1.3}
\end{eqnarray} 
The new Hamiltonian of the sample-tip system becomes 
\begin{equation}
K_T = K + V_t, \quad K = K_s + K_p , \label{eqn2.1.4}
\end{equation} 
where the tunneling interaction $V_t$ is defined by 
\begin{equation}
V_t = T_0\sum_{\sigma} (d^\dag_{j\sigma} c_{j\sigma} + c^\dag_{j\sigma} d_{j\sigma}) . \label{eqn2.1.5}
\end{equation}
Here we model the tunneling interaction $V_t$ with atomic resolution, where the sample electrons at location $j$ can tunnel through the vacuum barrier to the corresponding location $j$ of the tip, and vice versa. This approach is for the atomic resolution of STM \citep{TersoffHamannPRL1983,TersoffHamannPRB1985}. It employs a simplified representation where quantum tunneling occurs exclusively through an atomically confined channel between the sample and tip. Our treatment is very similar to two theoretical perspectives: (i) an electrode-atom-electrode model of the STM junction \citep{LangSTM1985}, and (ii) the identification of the tip apex atom as the dominant contributor to the sample-tip interaction \citep{ChenCJJVC1988,ChenCJPRL1990}. It should be noted that the location of the tunneling tip electrons may be represented using various notations. The choice of $j$ is made here solely for simplicity. The tunneling constant $T_0$ is taken to be independent of both spin and spatial location. Under finite bias voltage, the tunneling interaction $V_t$ gives rise to a finite quantum tunneling current across the vacuum barrier, as illustrated schematically in Fig.~\ref{fig2} (c). 

The tunneling current operator can be defined by $I = e \frac{d N_p}{d t}$, where $N_p$ is the number operator of the tip electrons. From the Heisenberg equation of motion, it can be shown that  
\begin{equation}
I = -\frac{i e T_0}{\hbar} \sum_{\sigma} (d^\dag_{j\sigma} c_{j\sigma} - c^\dag_{j\sigma} d_{j\sigma}) . \label{eqn2.1.6}
\end{equation}
For an arbitrary observable operator $O$, its expectation value at time $t > t_i$ is given by
\begin{equation}
\overline{\langle O(t) \rangle} \equiv \langle O_H(t) \rangle = \operatorname{Tr} [\rho_0 O_H(t)],  \label{eqn2.1.7}
\end{equation}
where $O_H(t)$ denotes the Heisenberg picture representation of $O(t)$, evolving according to
\begin{equation}
O_H(t) = U_H^\dagger(t, t_i) O(t) U_H(t, t_i).  \label{eqn2.1.8}
\end{equation}
The time-evolution operator $U_H(t, t_i)$ takes the form
\begin{equation}
U_H(t, t_i) = T_t \exp\left[-\frac{i}{\hbar} \int_{t_i}^t dt_1 K_T(t_1)\right], \label{eqn2.1.9}
\end{equation}
with $T_t$ being a time-ordering operator. The interaction picture representation $O_I(t)$ is analogously defined through
\begin{equation}
O_I(t) = U_0^\dagger(t, t_i) O(t) U_0(t, t_i),  \label{eqn2.1.10}
\end{equation}
where the time-evolution operator $U_0(t, t_i)$ is given by
\begin{equation}
U_0(t, t_i) = T_t \exp\left[-\frac{i}{\hbar} \int_{t_i}^t dt_1 K(t_1)\right]. \label{eqn2.1.11}
\end{equation}
Introducing the $S$ matrix operator as $S(t, t_i) \equiv U_0^\dagger(t, t_i) U_H(t, t_i)$, we have the following transformation relation
\begin{equation}
O_H(t) = S(t_i, t) O_I(t) S(t, t_i), \label{eqn2.1.12} 
\end{equation} 
where $ S(t_i, t)=[S(t, t_i)]^\dagger $. The explicit form of $S(t, t_i)$ can be shown as following: 
\begin{equation}
S(t, t_i) = T_t \exp\left[-\frac{i}{\hbar} \int_{t_i}^t dt_1 V_I(t_1)\right],  \label{eqn2.1.13}
\end{equation}
where $V_I(t)$ represents the tunneling interaction in the interaction picture. 
With these definitions, we can derive the expectation value of the tunneling current as 
\begin{equation}
\overline{\langle I(t) \rangle} = \langle S(t_i, t) I_I(t) S(t, t_i) \rangle , \label{eqn2.1.14}
\end{equation}
where $I_I(t)$ represents the tunneling current operator in the interaction picture and $\langle O \rangle \equiv \operatorname{Tr} [\rho_0 O]$ denotes the ensemble statistical average. The conservation of the tip electron number in the initial equilibrium state implies $\langle d_{j\sigma} \rangle = \langle d^\dag_{j\sigma} \rangle = 0$, 
which immediately gives $\langle I_I(t)\rangle =0$. Within the linear-response approximation, we can obtain the expression
\begin{equation}
\overline{\langle I(t)\rangle} = -\frac{i}{\hbar} \int_{t_i}^t d t_1 \langle [I_I(t), V_I(t_1)] \rangle . \label{eqn2.1.15}
\end{equation}
This result represents the well-established linear-response theory for the quantum tunneling current in the single-tip STS \citep{Mahan1990}.

For the case we discuss here, $K$ is time independent. Thus, the expectation value of any observable $\overline{\langle O(t)\rangle}$ is initial time $t_i$ independent. In this case, $O_I(t)$ has an equivalent expression 
\begin{equation}
O_I(t) = e^{\frac{i}{\hbar} K t} O(t) e^{-\frac{i}{\hbar} K t} . \label{eqn2.1.16}
\end{equation}
Since $[K_s^{(0)}+K_p^{(0)}, \sum_{i\sigma} d_{i\sigma}^\dag d_{i\sigma}] = 0$, $c_{j\sigma,I}(t)$ and $d_{j\sigma,I}(t)$ have the following expressions  
\begin{equation}
c_{j\sigma,I}(t) = c_{j\sigma}(t), \  d_{j\sigma,I}(t) = e^{\frac{i}{\hbar} e V t} d_{j\sigma}(t) , \label{eqn2.1.17}
\end{equation}
where $c_{j\sigma}(t)$ and $d_{j\sigma}(t)$ are defined as
\begin{eqnarray}
c_{j\sigma}(t) &=& e^{\frac{i}{\hbar} K_s^{(0)} t}  c_{j\sigma} e^{-\frac{i}{\hbar} K_s^{(0)} t} , \notag \\
d_{j\sigma}(t) &=& e^{\frac{i}{\hbar} K_p^{(0)} t}  d_{j\sigma} e^{-\frac{i}{\hbar} K_p^{(0)} t} . \label{eqn2.1.18} 
\end{eqnarray}
Introducing the notations $A^{(\pm)}$ as 
\begin{equation}
A^{(+)}(t) \equiv \sum_\sigma d^\dag_{j\sigma} (t) c_{j\sigma}(t) , \  A^{(-)}(t) \equiv  [A^{(+)}(t)]^\dag , \label{eqn2.1.19}
\end{equation}
$V_I$ and $I_I$ can be reexpressed as 
\begin{eqnarray}
V_I(t) &=& T_0 \sum_{a=\pm} A^{(a)} (t) e^{i a \phi(t)} , \notag \\
I_I(t) &=& -\frac{i e T_0}{\hbar} \sum_{a=\pm} a\cdot A^{(a)} (t) e^{i a \phi(t)} , \label{eqn2.1.20}
\end{eqnarray}   
where $\phi$ is defined by 
\begin{equation}
\phi(t) = -\frac{1}{\hbar} e V t . \label{eqn2.1.21}
\end{equation}
With these definitions, $\overline{\langle I(t)\rangle}$ can be show to follow
\begin{eqnarray}
\overline{\langle I(t)\rangle} = -\frac{i e T_0^2}{\hbar^2} \int_{-\infty}^{+\infty} d t_1 &&[\chi_R(t-t_1) e^{-\frac{i}{\hbar} e V (t-t_1)}  \notag \\
&& - \chi_A(t_1-t) e^{\frac{i}{\hbar} e V (t-t_1)}] , \label{eqn2.1.22}
\end{eqnarray}
where $\chi_R$ and $\chi_A$ are defined by
\begin{eqnarray}
\chi_R(t-t_1) &=& -i \theta(t-t_1)\langle [A^{(+)}(t), A^{(-)}(t_1)] \rangle , \notag \\
\chi_A(t-t_1) &=& i \theta(t_1-t)\langle [A^{(+)}(t), A^{(-)}(t_1)] \rangle . \label{eqn2.1.23}
\end{eqnarray}
Here we have set $t_i\rightarrow -\infty$ in Eq. (\ref{eqn2.1.22}). Since $\chi_A(t) = [\chi_R(-t)]^{*} $ and $\chi_A(\omega) = [\chi_R(\omega)]^{*}$, where $\chi_R(\omega)= \int_{-\infty}^{+\infty} d t \chi_R(t) e^{i\omega t}$ and $\chi_A(\omega)= \int_{-\infty}^{+\infty} d t \chi_A(t) e^{i\omega t}$, it can be shown that 
\begin{equation}
\overline{\langle I(t)\rangle} = \frac{2 e T_0^2}{\hbar^2} \operatorname{Im} \chi_R(-\frac{1}{\hbar} e V) . \label{eqn2.1.24}
\end{equation}

Let us introduce the imaginary-time Green's function defined as $\chi(\tau) = -\langle T_\tau A^{(+)}(\tau) A^{(-)} (0) \rangle$. Using a standard method \citep{Mahan1990, Colemanbook0}, $\chi_R(\omega)$ can be obtained through analytic continuation $\chi_R(\omega) = \hbar \, \chi(i\nu_n)\big|_{i\nu_n \rightarrow \hbar \omega + i \delta^+}$. It yields the explicit expression
\begin{eqnarray}
\chi_R(\omega)&=&\frac{\hbar}{(2\pi)^2}\sum_{\sigma}\iint d\varepsilon_1 d\varepsilon_2 A_\sigma^{(c)}(\varepsilon_1) A_\sigma^{(d)}(\varepsilon_2) \notag \\
&&\times \frac{n_F(\varepsilon_2)-n_F(\varepsilon_1)}{\hbar \omega + \varepsilon_2 -\varepsilon_1 + i\delta	^+} , \label{eqn2.1.25} 
\end{eqnarray}
where $A_\sigma^{(c)}(\varepsilon)$ and $A_\sigma^{(d)}(\varepsilon)$ denote the single-particle spectral functions of the sample and tip electrons, respectively, defined by 
\begin{eqnarray}
A_\sigma^{(c)}(\varepsilon) &=& - 2 \operatorname{Im} G^{(c)}_{j\sigma} (i\omega_n)\big|_{i\omega_n \rightarrow \varepsilon + i\delta^+} , \notag \\
A_\sigma^{(d)}(\varepsilon) &=& - 2 \operatorname{Im} G^{(d)}_{j\sigma} (i\omega_n)\big|_{i\omega_n \rightarrow \varepsilon + i\delta^+} . \label{eqn2.1.26} 
\end{eqnarray}
Here $G^{(c)}_{j\sigma}(i\omega_n) = \int_0^{\beta} d \tau G^{(c)}_{j\sigma}(\tau) e^{i\omega_n \tau} $ with $G^{(c)}_{j\sigma}(\tau)= -\langle T_\tau c_{j\sigma}(\tau) c_{j\sigma}^\dag(0) \rangle$ and $G^{(d)}_{j\sigma}(i\omega_n) = \int_0^{\beta} d\tau G^{(d)}_{j\sigma}(\tau) e^{i\omega_n \tau} $ with $G^{(d)}_{j\sigma}(\tau)= -\langle T_\tau d_{j\sigma}(\tau) d_{j\sigma}^\dag(0) \rangle$. From Eqs. (\ref{eqn2.1.24}) and (\ref{eqn2.1.25}), the expectation value of the tunneling current can be expressed as  
\begin{eqnarray}
\overline{\langle I(t)\rangle} &=& -\frac{e T_0^2}{2\pi\hbar} \sum_{\sigma} \int d\varepsilon A^{(c)}_\sigma(\varepsilon - e V) A^{(d)}_\sigma(\varepsilon) \notag \\
&& \times [n_F(\varepsilon)-n_F(\varepsilon - e V)] , \label{eqn2.1.27}
\end{eqnarray}
where $n_F(\varepsilon)$ is the Fermi-Dirac distribution function. 

Let us introduce a local dynamical conductance 
\begin{equation}
\sigma_j(V) = \frac{d }{d V} \overline{\langle I(t)\rangle} . \label{eqn2.1.28}
\end{equation}
In the low-temperature limit with the following approximations,
\begin{equation}
n_F(x)\simeq \theta (-x) ,\  \frac{d n_F(x)}{dx} \simeq -\delta(x) , \label{eqn2.1.29}
\end{equation} 
and assuming a nearly constant single-particle spectral function for tip electrons near the Fermi energy, the expression for $\sigma_j(V)$ simplifies to
\begin{equation}
\sigma_j(V) = \frac{e^2 T_0^2}{2\pi \hbar} \sum_\sigma A^{(c)}_\sigma(-e V) A^{(d)}_\sigma (0) . \label{eqn2.1.30}
\end{equation}
For typical experimental configurations where the tip consists of a metallic material, the spectral function of the tip electrons is spin independent, 
\begin{equation}
A^{(d)}_\sigma (\xi) = A_d (\xi) . \label{eqn2.1.31}
\end{equation}
It takes the form
\begin{equation}
A_d (\xi) = 2\pi \rho_d(\xi) ,\ \rho_d(\xi)\equiv \frac{1}{N}\sum_{\mathbf{q}}\delta(\xi -\xi_{\mathbf{q}}) , \label{eqn2.1.32}
\end{equation}
where $\rho_d (\xi)$ represents the density of states of the tip electrons with energy dispersion $\xi_{\mathbf{q}}$. In contrast, the spectral function of the sample electrons has the general expression
\begin{eqnarray}
A^{(c)}_\sigma (\varepsilon)&=& \frac{2\pi}{Z_s}\sum_{\alpha\beta} (e^{-\beta E_\alpha} + e^{-\beta E_\beta}) \langle \alpha| c_{j\sigma} |\beta\rangle \langle \beta| c^\dag_{j\sigma} |\alpha\rangle \notag \\
&& \times \delta(\varepsilon + E_\alpha - E_\beta) , \label{eqn2.1.33} 
\end{eqnarray}
where $Z_s$ denotes the partition function of the sample electrons, while $E_\alpha$ and $E_\beta$ are their eigenenergies. Eq. (\ref{eqn2.1.30}) reveals that the local dynamical conductance can probe directly the single-particle spectral function of the sample electrons, establishing the single-tip STS as a powerful technique for atomic-scale electronic structure characterization.

\subsection{Basic principle: Nonequilibrium theory} \label{sec2.2}

In this part, we will develop an alternative nonequilibrium theory for the single-tip STS, which can naturally generalize to the double-tip STS case. The theory begins with the contour-ordered $S$-matrix defined as $S_c(t_i, t_i) = T_c [S(t_i, t) S(t, t_i)]$, which follows
\begin{equation}
S_c(t_i, t_i) = T_c e^{-\frac{i}{\hbar} \int_c d t_1 V_I(t_1) } . \label{eqn2.2.1}
\end{equation}
Here the time contour runs from $t_i$ to $t$ and back to $t_i$ as $t_i\rightarrow t \rightarrow t_i$. Within this framework, the observable operator in the Heisenberg picture takes the compact form as 
\begin{equation}
O_H(t) = T_c [S_c (t_i, t_i) O_I(t)] . \label{eqn2.2.2}
\end{equation}
The tunneling current expectation value at time $t>t_i$ is then expressed as  
\begin{equation}
\overline{\langle I(t)\rangle} = \langle T_c [S_c (t_i, t_i) I_I(t)] \rangle . \label{eqn2.2.3}
\end{equation}
In the linear-response approximation, this simplifies to
\begin{equation}
\overline{\langle I(t)\rangle} = -\frac{i}{\hbar} \int_c d t_1 \langle T_c [I_I(t) V_I(t_1)] \rangle , \label{eqn2.2.4}
\end{equation}
where $V_I(t)$ and $I_I(t)$ are given in Eq. (\ref{eqn2.1.20}). Crucially, the conservation of the tip electron number leads to two distinct contributions to $\overline{\langle I(t)\rangle}$,   
\begin{equation}
\overline{\langle I(t)\rangle} = \overline{\langle I_{+}(t)\rangle} + \overline{\langle I_{-}(t)\rangle} , \label{eqn2.2.5}   
\end{equation}
with the explicit expressions of $\overline{\langle I_{+}(t)\rangle}$ and $\overline{\langle I_{-}(t)\rangle}$ given by 
\begin{eqnarray}
\overline{\langle I_{+}(t)\rangle} &=& e\left(\frac{-i T_0}{\hbar}\right)^2 \int_c d t_1 \langle T_c [A^{(+)}(t) A^{(-)}(t_1)]\rangle \notag \\
&& \times e^{i[\phi(t)-\phi(t_1)]} , \label{eqn2.2.6} \\
\overline{\langle I_{-}(t)\rangle} &=& - e\left(\frac{-i T_0}{\hbar}\right)^2 \int_c d t_1 \langle T_c [A^{(-)}(t) A^{(+)}(t_1)]\rangle \notag \\
&& \times e^{-i[\phi(t)-\phi(t_1)]} .  \label{eqn2.2.7}
\end{eqnarray}

Building upon the well-established nonequilibrium Green's function formalism \citep{Langreth1976, Rammer2007, HaugJauho2008}, we introduce the contour-ordered Green's functions 
\begin{eqnarray}
G_c(j\sigma; t_1, t) = -i\langle T_c c_{j\sigma}(t_1) c^\dag_{j\sigma} (t) \rangle , \label{eqn2.2.8} \\
G_d(j\sigma; t_1, t) = -i\langle T_c d_{j\sigma}(t_1) d^\dag_{j\sigma} (t) \rangle , \label{eqn2.2.9} 
\end{eqnarray}
which allows us to express the correlation function as 
\begin{equation}
\langle T_c [A^{(+)}(t) A^{(-)}(t_1) \rangle = G_{c}(j\sigma; t, t_1) G_d (j\sigma; t_1, t) . \label{eqn2.2.10}
\end{equation}
Consequently, $\overline{\langle I_{+}(t)\rangle}$ becomes as
\begin{eqnarray}
\overline{\langle I_{+}(t)\rangle} &=& -\frac{e T_0^2}{\hbar^2} \sum_\sigma \int_c d t_1 G_c(j\sigma; t, t_1) G_d(j\sigma; t_1, t) \notag \\
&& \times e^{\frac{i}{\hbar}e V (t_1 - t)} .   \label{eqn2.2.11}
\end{eqnarray}
Let us introduce the greater and lesser Green's functions
\begin{eqnarray}
G^{>}_{d}(j\sigma; t_1, t) &=& -i \langle d_{j\sigma}(t_1) d^{\dag}_{j\sigma}(t) \rangle , \notag \\
G^{<}_{d}(j\sigma; t_1, t) &=& i \langle d^{\dag}_{j\sigma}(t)  d_{j\sigma}(t_1) \rangle . \label{eqn2.2.12}
\end{eqnarray}
The contour-ordered Green's function can be reexpressed into the following form: 
\begin{eqnarray}
&& G_d(j\sigma; t_1, t) \notag \\
&=& \theta_c(t_1 - t) G^{>}_d(j\sigma; t_1, t) + \theta_c(t-t_1) G^{<}_d(j\sigma; t_1, t)  , \quad \label{eqn2.2.13}
\end{eqnarray}
where $\theta_c(t_1 -t)$ and $\theta_c(t-t_1)$ are step functions defined on the time contour.  The Fourier transformations of the greater and lesser Green's functions are defined as 
\begin{eqnarray}
G^{(l)}_d (j\sigma; \omega) &=& \int_{-\infty}^{+\infty} d t G^{(l)}_d (j\sigma; t) e^{i\omega t} , \notag\\
G^{(l)}_d (j\sigma; t) &=& \frac{1}{2\pi}\int_{-\infty}^{+\infty} d \omega G^{(l)}_d(j\sigma; \omega) e^{-i\omega t} , \label{eqn2.2.14}
\end{eqnarray}
where $l=>$ or $<$, and we have assumed the time-translation invariance $G^{(l)}_d (j\sigma; t_1, t)= G^{(l)}_d (j\sigma; t_1 - t)$. 
The expression of $\overline{\langle I_{+}(t)\rangle}$ then transforms to 
\begin{equation}
\overline{\langle I_{+}(t)\rangle} = -\frac{e T_0^2}{2\pi \hbar^2} \sum_{\sigma l}\int d \omega G^{(+)}_{c,l}(j\sigma;\omega-\frac{1}{\hbar} e V) G^{(l)}_d(j\sigma; \omega) , \label{eqn2.2.15}
\end{equation}
where $G^{(+)}_{c,l}$ are defined by 
\begin{equation}
G^{(+)}_{c,l}(j\sigma; \omega) = \int_c d t_1 \overline{\theta}^{(l)}_c(t_1 - t) G_c(j\sigma; t, t_1) e^{-i\omega (t_1 - t)} . \label{eqn2.2.16} 
\end{equation}
Here the contour step functions are defined as
\begin{equation}
\overline{\theta}^{(l)}_c (t) = \left\{ 
\begin{array}{ll}
\theta_c(t), & l=> , \\
\theta_c(-t). & l=< .
\end{array}  \right. \label{eqn2.2.17}
\end{equation}
For a metallic tip with nearly free electronic states, it can be easily shown that 
\begin{equation}
G^{(l)}_d (j\sigma; \omega) =\frac{2\pi i}{N}\sum_{\mathbf{q}} G^{(l)}_d (\xi_{\mathbf{q}}) \delta(\omega - \frac{1}{\hbar}\xi_{\mathbf{q}}) , \label{eqn2.2.18} 
\end{equation}
where $G^{(l)}_d$ are given by 
\begin{equation}
G^{>}_d (\xi_{\mathbf{q}}) = n_F(\xi_{\mathbf{q}})-1 , \, G^{<}_d (\xi_{\mathbf{q}}) = n_F(\xi_{\mathbf{q}}) . \label{eqn2.2.19}
\end{equation}
The final expression of $\overline{\langle I_{+}(t)\rangle}$ becomes 
\begin{eqnarray}
&& \overline{\langle I_{+}(t)\rangle} \notag \\
&=& -\frac{i e T_0^2}{2\pi\hbar^2} \sum_{\sigma l} \int d\xi G^{(+)}_{c,l}[j\sigma;\frac{1}{\hbar}(\xi- e V)] G^{(l)}_d(\xi) A_d(\xi) , \qquad \label{eqn2.2.20}
\end{eqnarray}
where $A_d(\xi)$ is the single-particle spectral function of the tip electrons defined in Eq. (\ref{eqn2.1.32}). 

Following an analogous derivation to that of $\overline{\langle I_{+}(t)\rangle}$, we can obtain the other tunneling current component 
\begin{eqnarray}
&& \overline{\langle I_{-}(t)\rangle} \notag \\ 
&=& \frac{i e T_0^2}{2\pi\hbar^2} \sum_{\sigma l} \int d\xi G^{(-)}_{c,l}[j\sigma;\frac{1}{\hbar}(\xi- e V)] G^{(l)}_d(\xi) A_d(\xi) , \qquad \label{eqn2.2.21}
\end{eqnarray}
where $G^{(-)}_{c,l}$ is defined by
\begin{equation}
G^{(-)}_{c,l}(j\sigma; \omega) = \int_c d t_1 \overline{\theta}^{(l)}_c(t - t_1) G_c(j\sigma; t_1, t) e^{i\omega (t_1 - t)} . \label{eqn2.2.22} 
\end{equation}

Let us define the composite Green's function
\begin{equation}
G_{c,l}(j\sigma; \omega) = G^{(+)}_{c,l}(j\sigma; \omega) - G^{(-)}_{c,l}(j\sigma; \omega) . \label{eqn2.2.23}
\end{equation}
It allows us to express the total tunneling current as
\begin{equation}
\overline{\langle I(t)\rangle} = -\frac{i e T_0^2}{2\pi\hbar^2} \sum_{\sigma l} \int d\xi G_{c,l}[j\sigma; \frac{1}{\hbar}(\xi- e V)] G^{(l)}_d(\xi) A_d(\xi) . \label{eqn2.2.24}
\end{equation}
Applying the low-temperature approximations defined by Eq. (\ref{eqn2.1.29}), we obtain
\begin{equation}
\frac{d}{d V} G^{(l)}_{d} (\xi + eV) = - e  \delta(\xi + eV). \label{eqn2.2.25}
\end{equation}
Furthermore, for the tip electrons near the Fermi energy, we make the additional approximation 
\begin{equation}
\frac{d}{d V} A_d(\xi + eV) = 0. \label{eqn2.2.26} 
\end{equation}
Combining these results, the local dynamical conductance can be shown to follow
\begin{equation}
\sigma_j(V) = \frac{i e^2 T_0^2}{2\pi \hbar^2} \sum_{\sigma l} G_{c,l}(j\sigma; -\frac{1}{\hbar}e V) A_d (0) . \label{eqn2.2.27}
\end{equation}

Let us introduce the following effective current operators
\begin{equation}
J_A (j\sigma; t, t_1; \omega) = J^{(+)}_A (j\sigma; t, t_1; \omega) + J^{(-)}_A (j\sigma; t, t_1; \omega) , \label{eqn2.2.28}
\end{equation}
where $J^{(+)}_A$ and $J^{(-)}_A$ are defined as
\begin{eqnarray}
J^{(+)}_A (j\sigma; t, t_1; \omega) &=& -i c_{j\sigma}(t) c^\dag_{j\sigma}(t_1) e^{-i\omega (t_1-t)} , \notag \\
J^{(-)}_A (j\sigma; t, t_1; \omega) &=& i c_{j\sigma}(t_1) c^\dag_{j\sigma}(t) e^{i\omega (t_1-t)} . \label{eqn2.2.29}
\end{eqnarray}
Next, let us define a first-order current correlation function 
\begin{equation}
G_J(j\sigma, \omega) = G^{(+)}_J(j\sigma, \omega) + G^{(-)}_J(j\sigma, \omega) , \label{eqn2.2.30}
\end{equation}
with $G^{(+)}_J$ and $G^{(-)}_J$ given by 
\begin{eqnarray}
G^{(+)}_J(j\sigma, \omega) &=& \int_c d t_1 \langle T_c J^{(+)}_A (j\sigma; t, t_1; \omega) \rangle , \notag \\
G^{(-)}_J(j\sigma, \omega) &=& \int_c d t_1 \langle T_c J^{(-)}_A (j\sigma; t, t_1; \omega) \rangle . \label{eqn2.2.31}
\end{eqnarray}
The following relations hold
\begin{eqnarray}
&&J^{(-)}_A (j\sigma; t, t_1; \omega) = [J^{(+)}_A (j\sigma; t, t_1; \omega)]^{\dag}, \label{eqn2.2.32}\\
&&G^{(-)}_J(j\sigma, \omega) = - [G^{(+)}_J(j\sigma, \omega)]^{*} . \label{eqn2.2.33}
\end{eqnarray}
Furthermore, it can be shown that
\begin{equation}
G_J(j\sigma, \omega) = \sum_{l}G_{c,l}(j\sigma; \omega) . \label{eqn2.2.34}
\end{equation}
Consequently, the local dynamical conductance can be rewritten as
\begin{equation}
\sigma_j(V) = \frac{i e^2 T_0^2}{2\pi \hbar^2} \sum_{\sigma }
G_J(j\sigma, -\frac{1}{\hbar}e V) A_d (0) . \label{eqn2.2.35}
\end{equation}
This is one main result for the single-tip STS we have obtained from the nonequilibrium theory. Eqs. (\ref{eqn2.2.30}) and (\ref{eqn2.2.33}) show that $\sigma_j(V)$ has another equivalent expression  
\begin{equation}
\sigma_j(V) = \frac{- e^2 T_0^2}{\pi \hbar^2} \sum_{\sigma } \operatorname{Im} G^{(+)}_J(j\sigma, -\frac{1}{\hbar}e V) A_d (0) . \label{eqn2.2.36}
\end{equation}
Additionally, it can be shown that 
\begin{equation}
-2 \operatorname{Im} G^{(+)}_J(j\sigma, \omega) = \hbar A^{(c)}_\sigma (\hbar \omega) . \label{eqn2.2.37}
\end{equation}
It allows us to recover the previous result we have obtained from equilibrium linear-response theory, Eq. (\ref{eqn2.1.30}).

\subsection{Fermi liquid state} \label{sec2.3}

Consider a system where the sample electrons are in a nearly free Fermi liquid state described by the Hamiltonian
\begin{equation}
K_s^{(0)} = \sum_{\mathbf{k}\sigma} \varepsilon_{\mathbf{k}} c^\dag_{\mathbf{k}\sigma} c_{\mathbf{k}\sigma} . \label{eqn2.3.1}
\end{equation}  
Here $c_{\mathbf{k}\sigma}$ is defined by
\begin{equation}
c_{\mathbf{k}\sigma} = \frac{1}{\sqrt{N}} \sum_{i} c_{i\sigma} e^{-i\mathbf{k}\cdot \mathbf{R}_{i}} . \label{eqn2.3.2}
\end{equation}
The single-particle spectral function $A^{(c)}_\sigma (\varepsilon)$ can be show to follow
\begin{equation}
A^{(c)}_\sigma (\varepsilon) = 2\pi \rho_c(\varepsilon) ,\ \rho_c(\varepsilon)\equiv \frac{1}{N}\sum_{\mathbf{k}}\delta(\varepsilon -\varepsilon_{\mathbf{k}}) , \label{eqn2.3.3}
\end{equation}
where $\rho_c(\varepsilon)$ denotes the density of states of the sample electrons. In this case, the local dynamical conductance reduces to
\begin{equation}
\sigma_j(V) = \frac{4 \pi e^2 T_0^2}{\hbar} \rho_c(-e V) \rho_d (0). \label{eqn2.3.4}
\end{equation}
Eq. (\ref{eqn2.3.4}) demonstrates that the local dynamical conductance can provide directly the local density of states of the sample electrons in Fermi liquid state.

\subsection{Superconducting state} \label{sec2.4}

Consider a superconductor where the sample electrons are in a spin-singlet superconducting state described by a BCS mean-field theory. The Hamiltonian is given by
\begin{equation}
K_s^{(0)} = \sum_{\mathbf{k}} \Psi^\dag_{\mathbf{k}} 
\left( \begin{array}{cc}
\varepsilon_{\mathbf{k}} & \Delta_{\mathbf{k}} \\
\Delta^{*}_\mathbf{k} & -\varepsilon_{\mathbf{k}}
\end{array} \right) \Psi_{\mathbf{k}} , \label{eqn2.4.1}
\end{equation}
where $\Delta_{\mathbf{k}}=|\Delta_{\mathbf{k}}|e^{i\theta_{\mathbf{k}}}$ represents the superconducting gap function and $\Psi_{\mathbf{k}}$ is the Nambu spinor operator defined as 
\begin{equation}
\Psi_{\mathbf{k}} = \left( \begin{array}{l}
c_{\mathbf{k}\uparrow} \\
c^\dag_{-\mathbf{k}\downarrow}
\end{array} \right) . \label{eqn2.4.2}
\end{equation}
Perform a Bogoliubov transformation 
\begin{equation}
\left( \begin{array}{l}
c_{\mathbf{k}\uparrow} \\
c^\dag_{-\mathbf{k}\downarrow}
\end{array} \right) =
U_{\mathbf{k}} \left( \begin{array}{l}
\alpha_{\mathbf{k}\uparrow} \\
\alpha^\dag_{-\mathbf{k}\downarrow}
\end{array} \right), \   
U_{\mathbf{k}}= \left( \begin{array}{cc}
\mu_\mathbf{k} & -\nu_{\mathbf{k}} \\
\nu^{*}_\mathbf{k} & \mu_{\mathbf{k}}
\end{array} \right) , \label{eqn2.4.3}
\end{equation}
where $\mu_{\mathbf{k}}= \sqrt{\frac{1}{2}(1+\frac{\varepsilon_{\mathbf{k}}}{E_{\mathbf{k}}})}$, $\nu_{\mathbf{k}}= e^{i\theta_{\mathbf{k}}} \sqrt{\frac{1}{2}(1-\frac{\varepsilon_{\mathbf{k}}}{E_{\mathbf{k}}})}$ with $E_{\mathbf{k}} = \sqrt{\varepsilon^2_{\mathbf{k}}+|\Delta_\mathbf{k}|^2}$. This transformation diagonalizes the BCS mean-field Hamiltonian to 
\begin{equation}
K_s^{(0)} = \sum_{\mathbf{k}} E_{\mathbf{k}} (\alpha^\dag_{\mathbf{k}\uparrow} \alpha_{\mathbf{k}\uparrow} + \alpha^\dag_{-\mathbf{k}\downarrow} \alpha_{-\mathbf{k}\downarrow} ) + E_0 , \label{eqn2.4.4}
\end{equation}
where $E_0 = -\sum_{\mathbf{k}} E_{\mathbf{k}}$. 

Following the standard procedures \citep{Mahan1990,Colemanbook0}, the single-particle spectral function can be obtained as 
\begin{equation}
A^{(c)}_\sigma(\varepsilon) = \frac{2\pi}{N}\sum_{\mathbf{k}} [\mu_{\mathbf{k}}^2 \delta(\varepsilon-E_{\mathbf{k}}) + |\nu_{\mathbf{k}}|^2 \delta(\varepsilon+E_{\mathbf{k}}) ] . \label{eqn2.4.5} 
\end{equation} 
The local dynamical conductance in the superconducting state then takes the form
\begin{eqnarray}
\sigma_j(V) &=& \frac{4\pi e^2 T_0^2}{N\hbar}\sum_{\mathbf{k}} [\mu_{\mathbf{k}}^2 \delta(e V + E_{\mathbf{k}}) + |\nu_{\mathbf{k}}|^2 \delta(e V - E_{\mathbf{k}}) ] \notag \\
&& \times \rho_d(0) .  \label{eqn2.4.6}
\end{eqnarray}
For the special case of $s$-wave pairing ($\Delta_{\mathbf{k}}=\Delta$), it simplifies to
\begin{equation}
\sigma_j(V)=  \theta(|eV|-\Delta) \frac{4\pi e^2 T_0^2 }{\hbar}  \rho_{c}(0) \rho_d(0) \frac{|eV|}{\sqrt{(eV)^2 - \Delta^2}} , \label{eqn2.4.7}
\end{equation}
where we have assumed that the density of states of the sample electrons remains approximately constant near the Fermi energy in the normal state.

\section{Coincidence double-tip scanning tunneling spectroscopy} \label{sec3}

In this section, we develop a new scanning tunneling spectroscopy technique, a coincidence double-tip STS. This technique extends conventional single-tip STS by enabling coincidence measurement of two spatially resolved tunneling currents. The resulting coincidence dynamical conductance can provide direct access to the spatially resolved two-body correlations of sample electrons.

\subsection{Basic principle} \label{sec3.1}

The coincidence double-tip STS can be implemented using the double-tip STM setup depicted schematically in Fig.~\ref{fig1} (b). In contrast to conventional single-tip STM, this configuration utilizes two independently biased probe tips, each producing a quantum tunneling current. The two tunneling currents are measured in coincidence, quantified by the correlation function $\overline{\langle I_1(t) I_2(t)\rangle}$. Below, we develop a nonequilibrium theory for the coincidence tunneling currents in the double-tip STS. The formalism constitutes a systematic second-order extension of single-tip STS theory, which naturally incorporates the two-body correlations of sample electrons.

Suppose prior to time $t_i$, the sample-tip composite system resides in an unbiased equilibrium state described by the density matrix
\begin{equation}
\rho_0 = \frac{1}{Z} e^{-\beta [K_s^{(0)} + K_{p_1}^{(0)} + K_{p_2}^{(0)}] } , \label{eqn3.1.1}
\end{equation}
where the partition function $Z=\operatorname{Tr} \{ \exp[-\beta (K_s^{(0)} + K_{p_1}^{(0)} + K_{p_2}^{(0)})]\}$ normalizes the equilibrium distribution, and the three Hamiltonians are defined by
\begin{eqnarray}
K_s^{(0)} &=& H_s -\mu_s \sum_{i\sigma} c^{\dag}_{i\sigma} c_{i\sigma}, \notag  \\
K_{p_1}^{(0)} &=& H_{p_1} -\mu_{p_1} \sum_{i\sigma} d^{\dag}_{1,i\sigma} d_{1,i\sigma}, \label{eqn3.1.2} \\
K_{p_2}^{(0)} &=& H_{p_2} -\mu_{p_2} \sum_{i\sigma} d^{\dag}_{2,i\sigma} d_{2,i\sigma}. \notag  
\end{eqnarray}
Here $d^{\dag}_{1,i\sigma} (d_{1,i\sigma})$ and $d^{\dag}_{2,i\sigma} (d_{2,i\sigma})$ are the creation (annihilation) operators for the tip-1 and tip-2 electrons, respectively, with $\mu_{p_1}$ and $\mu_{p_2}$ representing the corresponding chemical potentials. 

At time $t_i$, two bias voltages $V_1$ and $V_2$ are applied, inducing two finite quantum tunneling currents. The subsequent dynamical evolution of the composite system is governed by the total Hamiltonian 
\begin{equation}
K_T = K + V_t , \label{eqn3.1.3}
\end{equation}
where 
\begin{equation}
K = K_s + K_{p_1} + K_{p_2}  \label{eqn3.1.4}
\end{equation}
incorporates the modified Hamiltonians 
\begin{eqnarray}
&& K_s = K_s^{(0)} , \notag \\
&& K_{p_1} = K_{p_1}^{(0)} - e V_1\sum_{i\sigma} d^{\dag}_{1,i\sigma} d_{1,i\sigma} ,  \label{eqn3.1.5} \\
&& K_{p_2} = K_{p_2}^{(0)} - e V_2\sum_{i\sigma} d^{\dag}_{2,i\sigma} d_{2,i\sigma} ,  \notag
\end{eqnarray}
and the tunneling Hamiltonian $V_t$ 
\begin{equation}
V_t = V_{t_1} + V_{t_2}  \label{eqn3.1.6}
\end{equation}
involves two tunneling interactions given by
\begin{eqnarray}
V_{t_1} &=& T_1\sum_{\sigma} (d^\dag_{1\sigma} c_{j_1\sigma} + c^\dag_{j_1\sigma} d_{1\sigma}) , \notag \\
V_{t_2} &=& T_2\sum_{\sigma} (d^\dag_{2\sigma} c_{j_2\sigma} + c^\dag_{j_2\sigma} d_{2\sigma}) . \label{eqn3.1.7}
\end{eqnarray}
Here, the tunneling interactions are defined to originate from two spatially localized processes, with electrons tunneling between the sample at distinct locations $j_1$ and $j_2$ and their respective tip positions. For conciseness, we have adopted the following simplified notations for the tip electron operators 
\begin{equation}
d_{1\sigma} \equiv d_{1,j_1\sigma} , \, d_{2\sigma} \equiv d_{2,j_2\sigma} , \label{eqn3.1.8}
\end{equation}
with the creation operators $d^\dag_{1\sigma}$ and $d^\dag_{2\sigma}$ defined analogously. The two corresponding tunneling current operators are defined by $I_1=e\frac{d N_{p_1}}{d t}$ and $I_2=e\frac{d N_{p_2}}{d t}$, which take the forms
\begin{eqnarray}
I_1 &=& -\frac{i e T_1}{\hbar} \sum_{\sigma} (d^\dag_{1\sigma} c_{j_1\sigma} - c^\dag_{j_1\sigma} d_{1\sigma}) , \notag \\
I_2 &=& -\frac{i e T_2}{\hbar} \sum_{\sigma} (d^\dag_{2\sigma} c_{j_2\sigma} - c^\dag_{j_2\sigma} d_{2\sigma}) . \label{eqn3.1.9}
\end{eqnarray}

At observable time $t > t_i$, the two tunneling currents are measured in coincidence. This can be descried by the coincidence tunneling currents defined by  
\begin{equation}
\overline{\langle I_1(t) I_2(t)\rangle} = \langle [I_{1H}(t) I_{2H}(t)] \rangle , \label{eqn3.1.10}
\end{equation}
where $I_{1H}(t)$ and $I_{2H}(t)$ denote the tunneling current operators in the Heisenberg picture defined following Eq. (\ref{eqn2.2.2}). The commutation $[I_1(t), I_2(t)]=0$ ensures the reality of the observable value of $\overline{\langle I_1(t) I_2(t)\rangle}$ because
\begin{equation}
[I_{1H}(t) I_{2H}(t)]^\dag = [I_{1H}(t) I_{2H}(t)] . \label{eqn3.1.11}
\end{equation}
The coincidence tunneling currents have a contour-ordered representation 
\begin{equation}
\overline{\langle I_1(t) I_2(t)\rangle} = \langle T_c S_c(t_i, t_i) I_{1I}(t) I_{2I}(t)  \rangle , \label{eqn3.1.12}
\end{equation}
where $S_c$ follows Eq. (\ref{eqn2.2.1}) with the modified tunneling Hamiltonian in the interaction picture 
\begin{equation}
V_I(t) = V_{1I}(t) + V_{2I}(t) . \label{eqn3.1.13}
\end{equation}
The two individual tunneling interactions in the interaction picture are given by 
\begin{eqnarray}
V_{1I}(t) &=& T_1 \sum_{a=\pm} A_1^{(a)} (t) e^{i a \phi_1(t)} , \notag \\
V_{2I}(t) &=& T_2 \sum_{a=\pm} A_2^{(a)} (t) e^{i a \phi_2(t)}  , \label{eqn3.1.14}
\end{eqnarray}   
with the fermionic bilinears $A_1^{(a)}$ and  $A_2^{(a)}$ $(a=\pm)$ defined as
\begin{eqnarray}
A_1^{(+)}(t) &\equiv & \sum_\sigma d^\dag_{1\sigma} (t) c_{j_1\sigma}(t) , \  A_1^{(-)}(t) \equiv  [A_1^{(+)}(t)]^\dag , \notag \\
A_2^{(+)}(t) &\equiv & \sum_\sigma d^\dag_{2\sigma} (t) c_{j_2\sigma}(t) , \  A_2^{(-)}(t) \equiv  [A_2^{(+)}(t)]^\dag . \quad \qquad \label{eqn3.1.15}
\end{eqnarray}
Here the time-dependent operators evolve as
\begin{eqnarray}
c_{j\sigma}(t) &=& e^{\frac{i}{\hbar} K_s^{(0)} t}  c_{j\sigma} e^{-\frac{i}{\hbar} K_s^{(0)} t} , \notag \\
d_{1\sigma}(t) &=& e^{\frac{i}{\hbar} K_{p_1}^{(0)} t}  d_{1\sigma} e^{-\frac{i}{\hbar} K_{p_1}^{(0)} t} , \label{eqn3.1.16} \\
d_{2\sigma}(t) &=& e^{\frac{i}{\hbar} K_{p_2}^{(0)} t}  d_{2\sigma} e^{-\frac{i}{\hbar} K_{p_2}^{(0)} t} , \notag
\end{eqnarray}
with similar time evolutions of $c^\dag_{j\sigma}(t)$, $d^\dag_{1\sigma}(t)$ and $d^\dag_{2\sigma}(t)$. The phase factors $\phi_1$ and $\phi_2$ are defined by
\begin{equation}
\phi_1(t) = -\frac{1}{\hbar} e V_1 t, \, \phi_2(t) = -\frac{1}{\hbar} e V_2 t . \label{eqn3.1.17}
\end{equation}
The tunneling current operators in the interaction picture are similarly given by
\begin{eqnarray}
I_{1I}(t) &=& -\frac{i e T_1}{\hbar} \sum_{a=\pm} a\cdot A_1^{(a)} (t) e^{i a \phi_1(t)}  , \notag \\
I_{2I}(t) &=& -\frac{i e T_2}{\hbar} \sum_{a=\pm} a\cdot A_2^{(a)} (t) e^{i a \phi_2(t)} . \label{eqn3.1.18}
\end{eqnarray}   

By expanding the $S_c$ matrix in a Taylor series, the perturbation contributions to the coincidence tunneling currents can be calculated as follows:
\begin{eqnarray}
&& \overline{\langle I_1(t) I_2(t)\rangle} \notag \\
&=& \langle I_{1I}(t) I_{2I}(t)\rangle \notag \\
&+& \left(-\frac{i}{\hbar}\right) \int_c d t_1 \langle T_c V_I(t_1) I_{1I}(t) I_{2I}(t) \rangle \notag \\
&+& \frac{1}{2} \left(\frac{-i}{\hbar} \right)^2 \iint_c d t_2 d t_1 \langle T_c V_I(t_2) V_I(t_1) I_{1I}(t) I_{2I}(t) \rangle .  \notag \\
&+& \cdots  . \label{eqn3.1.19} 
\end{eqnarray}
Due to the conservation of the electron numbers in the two tips, we have $\langle d_{1\sigma} \rangle = \langle d^\dag_{1\sigma} \rangle = 0$ and $\langle d_{2\sigma} \rangle = \langle d^\dag_{2\sigma} \rangle = 0$.  Consequently, both the zeroth- and first-order contributions to the coincidence tunneling currents vanish, {\it i.e.},
\begin{equation}
\langle I_{1I}(t) I_{2I}(t)\rangle = 0 , \label{eqn3.1.20}
\end{equation}
and 
\begin{equation}
\langle T_c V_I(t_1) I_{1I}(t) I_{2I}(t) \rangle  = 0 . \label{eqn3.1.21}
\end{equation}
Thus, the lowest-order non-vanishing contribution arises from the second-order expansion of the $S_c$ matrix, 
\begin{eqnarray}
&& \overline{\langle I_1(t) I_2(t)\rangle} \notag \\
&=& \frac{1}{2} \left(\frac{-i}{\hbar} \right)^2 \iint_c d t_2 d t_1 \langle T_c V_I(t_2) V_I(t_1) I_{1I}(t) I_{2I}(t) \rangle .  \notag \\
&&   \label{eqn3.1.22}
\end{eqnarray} 
It can be further shown that $\overline{\langle I_1(t) I_2(t)\rangle}$ can decompose into a sum over several contributions
\begin{equation}
\overline{\langle I_1(t) I_2(t)\rangle} = \sum_{a_1 a_2 = \pm} \overline{\langle I_1(t) I_2(t)\rangle}^{(a_1 a_2)} , \label{eqn3.1.23}
\end{equation}
where these contributions take the explicit forms as
\begin{eqnarray}
&&\overline{\langle I_1(t) I_2(t)\rangle}^{(a_1 a_2)}  \notag \\
&=& \frac{(e T_1 T_2)^2}{\hbar^4} \iint_c d t_2 d t_1 e^{i a_1 [\phi_1(t)-\phi_1(t_1)]+i a_2[\phi_2(t)-\phi_2(t_2)]} \notag \\
&& \times a_1 a_2 \cdot \langle T_c A_1^{(a_1)}(t) A_2^{(a_2)}(t) A_2^{(\overline{a}_2)}(t_2) A_1^{(\overline{a}_1)}(t_1) \rangle.  \label{eqn3.1.24}
\end{eqnarray}
Here we have introduced the notations 
\begin{equation}
\overline{a}_1 \equiv - a_1, \, \overline{a}_2 \equiv - a_2 . \label{eqn3.1.25}
\end{equation}

Let us begin by considering $\overline{\langle I_1(t) I_2(t)\rangle}^{(++)}$. To facilitate our analysis, we introduce the following contour-ordered Green's functions for the sample electrons   
\begin{eqnarray}
&& G_c(j_1\sigma_1 t_1, j_2\sigma_2 t_2; j_2\sigma_2 t_2^\prime, j_1\sigma_1 t_1^\prime) \notag \\
&=& (-i)^2 \langle T_c c_{j_1\sigma_1}(t_1) c_{j_2\sigma_2}(t_2) c^\dag_{j_2\sigma_2}(t_2^\prime) c^\dag_{j_1\sigma_1} (t_1^\prime) \rangle, \qquad  \label{eqn3.1.26}
\end{eqnarray}
and for the tip electrons 
\begin{eqnarray}
G_{d_1}(j_1\sigma_1;  t_1, t) &=& -i \langle T_c d_{1\sigma_1}(t_1) d^\dag_{1\sigma_1} (t) \rangle , \notag \\
G_{d_2}(j_2\sigma_2;  t_2, t) &=& -i \langle T_c d_{2\sigma_2}(t_2) d^\dag_{2\sigma_2} (t) \rangle . \label{eqn3.1.27}
\end{eqnarray}
With these definitions, we derive the following expression as 
\begin{eqnarray}
&&\langle T_c A_1^{(+)}(t) A_2^{(+)}(t) A_2^{(-)}(t_2) A_1^{(-)}(t_1) \rangle \notag \\
&=& \sum_{\sigma_1 \sigma_2} G_c(j_1\sigma_1 t, j_2\sigma_2 t; j_2\sigma_2 t_2, j_1\sigma_1 t_1) \cdot G_{d_1}(j_1\sigma_1; t_1, t) \notag \\
&& \times G_{d_2}(j_2\sigma_2; t_2, t) . \label{eqn3.1.28}
\end{eqnarray}
Following the methodology established in the nonequilibrium theory for the single-tip STS, we also introduce the greater and lesser Green's functions for the tip electrons:
\begin{eqnarray}
G^{>}_{d_1}(j_1\sigma_1; t_1, t) &=& -i \langle d_{1\sigma_1}(t_1) d^{\dag}_{1\sigma_1}(t) \rangle , \notag \\
G^{<}_{d_1}(j_1\sigma_1; t_1, t) &=& i \langle d^{\dag}_{1\sigma_1}(t)  d_{1\sigma_1}(t_1) \rangle , \label{eqn3.1.29} 
\end{eqnarray}
and 
\begin{eqnarray}
G^{>}_{d_2}(j_2\sigma_2; t_2, t) &=& -i \langle d_{2\sigma_2}(t_2) d^{\dag}_{2\sigma_2}(t) \rangle , \notag \\
G^{<}_{d_2}(j_2\sigma_2; t_2, t) &=& i \langle d^{\dag}_{2\sigma_2}(t)  d_{2\sigma_2}(t_2) \rangle . \label{eqn3.1.30} 
\end{eqnarray}
The contour-ordered Green's functions for the tip electrons can then be expressed as
\begin{eqnarray}
G_{d_1}(j_1\sigma_1; t_1, t) &=& \theta_c(t_1 - t) G^{>}_{d_1}(j_1\sigma_1; t_1, t) \notag \\
&& + \theta_c(t-t_1) G^{<}_{d_1}(j_1\sigma_1; t_1, t) , \label{eqn3.1.31} \\
G_{d_2}(j_2\sigma_2; t_2, t) &=& \theta_c(t_2 - t) G^{>}_{d_2}(j_2\sigma_2; t_2, t) \notag \\
&& + \theta_c(t-t_2) G^{<}_{d_2}(j_2\sigma_2; t_2, t) . \label{eqn3.1.32}  
\end{eqnarray}
In analogy with Eq. (\ref{eqn2.2.14}), we further introduce the Fourier transformations of the greater and lesser Green's functions $G^{(l)}_{d_1}$ and $G^{(l)}_{d_2}$ (where $l=>$ or $<$). This allows us to express $\overline{\langle I_1(t) I_2(t)\rangle}^{(++)}$ as
\begin{widetext}
\begin{equation}
\overline{\langle I_1(t) I_2(t)\rangle}^{(++)} = \left(\frac{e T_1 T_2}{2\pi \hbar^2}\right)^2 \sum_{\sigma_1 \sigma_2 l_1 l_2} \iint d\omega_1 d\omega_2 G^{(++)}_{c,l_1 l_2}(j_1\sigma_1, j_2\sigma_2; \omega_1 - \frac{1}{\hbar}eV_1, \omega_2 - \frac{1}{\hbar} e V_2) G^{(l_1)}_{d_1}(j_1\sigma_1,\omega_1) G^{(l_2)}_{d_2}(j_2\sigma_2,\omega_2) , \label{eqn3.1.33}
\end{equation}
where $G^{(++)}_{c,l_1 l_2}$ is defined by
\begin{equation}
G^{(++)}_{c,l_1 l_2}(j_1\sigma_1, j_2\sigma_2; \omega_1, \omega_2) = \iint_c d t_2 d t_1 \overline{\theta}^{(l_1)}_c(t_1-t) \overline{\theta}^{(l_2)}_c(t_2-t) G_c(j_1\sigma_1 t, j_2\sigma_2 t; j_2\sigma_2 t_2, j_1\sigma_1 t_1) e^{-i\omega_1(t_1-t)-i\omega_2 (t_2-t)} . \label{eqn3.1.34} 
\end{equation}
Here, $\overline{\theta}^{(l)}_c(t)$ follows the definition given in Eq. (\ref{eqn2.2.17}). The functions $G^{(l_1)}_{d_1}$ and $G^{(l_2)}_{d_2}$ in Eq. (\ref{eqn3.1.33}) adopt forms analogous to Eq. (\ref{eqn2.2.18}),
\begin{equation}
G^{(l_1)}_{d_1} (j_1\sigma_1; \omega_1) =\frac{2\pi i}{N}\sum_{\mathbf{q}} G^{(l)}_{d_1} [\xi^{(1)}_{\mathbf{q}}] \, \delta[\omega_1 - \frac{1}{\hbar}\xi^{(1)}_{\mathbf{q}}] , \ G^{(l_2)}_{d_2} (j_2\sigma_2; \omega_2) =\frac{2\pi i}{N}\sum_{\mathbf{q}} G^{(l_2)}_{d_2} [\xi^{(2)}_{\mathbf{q}}] \, \delta[\omega_2 - \frac{1}{\hbar}\xi^{(2)}_{\mathbf{q}}] .  \label{eqn3.1.35} 
\end{equation}
Next, we define the spectral functions for the tip electrons
\begin{equation}
A_{d_1}(\xi_1) = 2\pi \rho_{d_1} (\xi_1) = \frac{2\pi}{N} \sum_{\mathbf{q}} \delta[\xi_1 - \xi^{(1)}_{\mathbf{q}}] , \ A_{d_2}(\xi_2) = 2\pi \rho_{d_2} (\xi_2) = \frac{2\pi}{N} \sum_{\mathbf{q}} \delta[\xi_2 - \xi^{(2)}_{\mathbf{q}}] , \label{eqn3.1.36}
\end{equation}
where $\xi^{(1)}_{\mathbf{q}}$ and $\xi^{(2)}_{\mathbf{q}}$
denote the energy dispersions of the tip-1 and tip-2 electrons, respectively. Using these definitions, $\overline{\langle I_1(t) I_2(t)\rangle}^{(++)}$ can be reexpressed into the following form: 
\begin{eqnarray}
\overline{\langle I_1(t) I_2(t)\rangle}^{(++)} &=& -\left(\frac{e T_1 T_2}{2\pi \hbar^2}\right)^2 \sum_{\sigma_1 \sigma_2 l_1 l_2} \iint d\xi_1 d\xi_2 G^{(++)}_{c,l_1 l_2}(j_1\sigma_1, j_2\sigma_2; \frac{1}{\hbar}(\xi_1 - eV_1), \frac{1}{\hbar}(\xi_2 - eV_2))  G^{(l_1)}_{d_1}(\xi_1) G^{(l_2)}_{d_2}(\xi_2) \notag \\
&& \times A_{d_1}(\xi_1) A_{d_2}(\xi_2) . \label{eqn3.1.37}
\end{eqnarray}
The remaining three contributions can be derived in a similar manner. Their detailed forms are provided in Appendix \ref{seca1}.

We define the generalized correlation function $G_{c,l_1 l_2}$ as 
\begin{equation}
G_{c,l_1 l_2} (j_1\sigma_1, j_2\sigma_2; \omega_1, \omega_2) = -\sum_{a_1 a_2 =\pm} (-1)^{a_1\cdot a_2} G_{c,l_1 l_2}^{(a_1 a_2)} (j_1\sigma_1, j_2\sigma_2; \omega_1, \omega_2) . \label{eqn3.1.38}
\end{equation}
Using this definition, $\overline{\langle I_1(t) I_2(t)\rangle}$ can be shown to follow
\begin{equation}
\overline{\langle I_1(t) I_2(t)\rangle} = \left(\frac{e T_1 T_2}{2\pi \hbar^2}\right)^2 \sum_{\sigma_1 \sigma_2 l_1 l_2} \iint d\xi_1 d\xi_2  G_{c,l_1 l_2}(j_1\sigma_1, j_2\sigma_2; \frac{1}{\hbar}(\xi_1 - eV_1), \frac{1}{\hbar}(\xi_2 - eV_2)) G^{(l_1)}_{d_1}(\xi_1) G^{(l_2)}_{d_2}(\xi_2) A_{d_1}(\xi_1) A_{d_2}(\xi_2) . \label{eqn3.1.39}
\end{equation}
This represents the fundamental result for the coincidence tunneling currents in our proposed double-tip STS. 
\end{widetext}

Let us introduce a coincidence dynamical conductance for the double-tip STS: 
\begin{equation}
\sigma_{j_1 j_2}(V_1, V_2) = \frac{d}{d V_2} \frac{d}{d V_1} \overline{\langle I_1(t) I_2(t)\rangle} .  \label{eqn3.1.40}
\end{equation}
Following the treatment in Eq. (\ref{eqn2.2.25}), we consider the low-temperature limit where the tip electron Green's function derivatives become 
\begin{eqnarray}
\frac{d}{d V_1} G^{(l_1)}_{d_1} (\xi_1 + eV_1) = - e \delta(\xi_1 + eV_1), \notag \\
\frac{d}{d V_2} G^{(l_2)}_{d_2} (\xi_2 + eV_2) = - e \delta(\xi_2 + eV_2) . \label{eqn3.1.41}
\end{eqnarray}
Furthermore, we assume the tip electron spectral functions remain constant near the Fermi level, 
\begin{equation}
\frac{d}{d V_1} A_{d_1}(\xi_1 + eV_1) = 0 , \, \frac{d}{d V_2} A_{d_2}(\xi_2 + eV_2) = 0. \label{eqn3.1.42} 
\end{equation}
Under these approximations, the coincidence dynamical conductance follows
\begin{eqnarray}
\sigma_{j_1 j_2}(V_1, V_2) &=&  \left(\frac{e^2 T_1 T_2}{2\pi \hbar^2}\right)^2 \sum_{\sigma_1 \sigma_2 l_1 l_2} G_{c,l_1 l_2}(j_1\sigma_1, j_2\sigma_2; \overline{\omega}_1, \overline{\omega}_2 ) \notag\\ && \times A_{d_1}(0) A_{d_2}(0),  \label{eqn3.1.43}
\end{eqnarray}
where the effective frequencies $\overline{\omega}_1$ and $\overline{\omega}_2$ are defined as
\begin{equation}
\overline{\omega}_1 = -\frac{1}{\hbar}eV_1 , \, \overline{\omega}_2 = -\frac{1}{\hbar}eV_2 . \label{eqn3.1.44}
\end{equation}

Building on the approach in Eq. (\ref{eqn2.2.28}), we introduce the following effective current operators 
\begin{eqnarray}
&&J_{1A} (j_1\sigma_1; t, t_1; \omega_1) \notag \\
&=& J^{(+)}_{1A} (j_1\sigma_1; t, t_1; \omega_1) + J^{(-)}_{1A} (j_1\sigma_1; t, t_1; \omega_1) , \notag \\
&& J_{2A} (j_2\sigma_2; t, t_2; \omega_2) \notag \\
&=& J^{(+)}_{2A} (j_2\sigma_2; t, t_2; \omega_2) + J^{(-)}_{2A} (j_2\sigma_2; t, t_2; \omega_2) , \label{eqn3.1.45}
\end{eqnarray}
where the constituent current operators $J^{(+)}_{1A}$ and $J^{(-)}_{1A}$ for the tip-1 electrons are given by
\begin{eqnarray}
J^{(+)}_{1A} (j_1\sigma_1; t, t_1; \omega_1) &=& -i c_{j_1\sigma_1}(t) c^\dag_{j_1\sigma_1}(t_1) e^{-i\omega_1 (t_1-t)} , \notag \\
J^{(-)}_{1A} (j_1\sigma_1; t, t_1; \omega_1) &=& i c_{j_1\sigma_1}(t_1) c^\dag_{j_1\sigma_1}(t) e^{i\omega_1 (t_1-t)} , \label{eqn3.1.46} 
\end{eqnarray}
and similarly for the tip-2 electron current operators $J^{(+)}_{2A}$ and $J^{(-)}_{2A}$ are defined by 
\begin{eqnarray}
J^{(+)}_{2A} (j_2\sigma_2; t, t_2; \omega_2) &=& -i c_{j_2\sigma_2}(t) c^\dag_{j_2\sigma_2}(t_2) e^{-i\omega_2 (t_2-t)} , \notag \\
J^{(-)}_{2A} (j_2\sigma_2; t, t_2; \omega_2) &=& i c_{j_2\sigma_2}(t_2) c^\dag_{j_2\sigma_2}(t) e^{i\omega_2 (t_2-t)} . \label{eqn3.1.47}
\end{eqnarray}
Let us define the second-order current correlation function
\begin{eqnarray}
&&G^{(2)}_{J}(j_1\sigma_1,j_2\sigma_2; \omega_1, \omega_2)  \notag \\
&=& \iint_c d t_2 d t_1 \langle T_c J_{1A} (j_1\sigma_1; t, t_1; \omega_1) J_{2A} (j_2\sigma_2; t, t_2; \omega_2) \rangle . \notag \\
&& \label{eqn3.1.48}
\end{eqnarray}
Remarkably, this connects to our earlier expression through
\begin{equation}
G^{(2)}_{J}(j_1\sigma_1,j_2\sigma_2; \omega_1, \omega_2)  = - \sum_{l_1 l_2} G_{c,l_1 l_2}(j_1\sigma_1,j_2\sigma_2; \omega_1, \omega_2) .  \label{eqn3.1.49}
\end{equation}
This relation reveals the deep connection between the current fluctuations and the sample electron correlations. Thus, the coincidence dynamical conductance takes its final form 
\begin{eqnarray}
\sigma_{j_1 j_2}(V_1, V_2) &=& -\left(\frac{e^2 T_1 T_2}{2\pi \hbar^2}\right)^2 \sum_{\sigma_1 \sigma_2} G^{(2)}_{J}(j_1\sigma_1, j_2\sigma_2; \overline{\omega}_1, \overline{\omega}_2 ) \notag \\
&& \times A_{d_1}(0) A_{d_2}(0),  \label{eqn3.1.50}
\end{eqnarray}
with $\overline{\omega}_1$ and $\overline{\omega}_2$ defined in Eq. (\ref{eqn3.1.44}). This represents a cornerstone result in our double-tip STS theory, demonstrating that the coincidence dynamical conductance directly probes the two-body correlations of sample electrons through the second-order current correlation function. Therefore, the double-tip STS provides a powerful technique for extracting the sample electron correlation features that are inaccessible in conventional single-tip STS measurements.

\subsection{Fermi liquid state} \label{sec3.2}

Let us study the double-tip STS of the sample electrons in a nearly free Fermi liquid state, which is described by the Hamiltonian in Eq. (\ref{eqn2.3.1}). Starting from the coincidence dynamical conductance given in Eq. (\ref{eqn3.1.50}), our first objective is to calculate the second-order current correlation function $G^{(2)}_{J}$. 
According to Wick's theorem \citep{Langreth1976,Rammer2007,HaugJauho2008}, 
$G^{(2)}_{J}$ consists of two distinct contributions
\begin{equation}
G^{(2)}_{J} = G^{(2)}_{J,D} + G^{(2)}_{J,E} , \label{eqn3.2.1}
\end{equation}
where the direct term $G^{(2)}_{J,D}$ is given by 
\begin{equation}
G^{(2)}_{J,D}(j_1\sigma_1,j_2\sigma_2; \omega_1, \omega_2) = G_{J}(j_1\sigma_1,\omega_1) \cdot G_{J}(j_2\sigma_2,\omega_2) , \label{eqn3.2.2} 
\end{equation}
and the exchange term $G^{(2)}_{J,E}$ will be derived in detail below. Correspondingly, the coincidence dynamical conductance also decomposes into two contributions
\begin{equation}
\sigma_{j_1 j_2}(V_1, V_2) = \sigma^{(D)}_{j_1 j_2}(V_1, V_2) + \sigma^{(E)}_{j_1 j_2}(V_1, V_2) . \label{eqn3.2.3}
\end{equation}
From Eq. (\ref{eqn3.2.2}), the direct conductance term $\sigma^{(D)}_{j_1 j_2}$ takes the form
\begin{equation}
\sigma^{(D)}_{j_1 j_2}(V_1, V_2) = \sigma_{j_1}(V_1) \cdot \sigma_{j_2}(V_2) , \label{eqn3.2.4}
\end{equation}
where $\sigma_{j_1}$ and $\sigma_{j_2}$ are given by
\begin{eqnarray}
\sigma_{j_1}(V_1) &=& \frac{i e^2 T_1^2}{2\pi \hbar^2} \sum_{\sigma_1 }
G_J(j_1\sigma_1, -\frac{1}{\hbar}e V_1) A_{d_1} (0), \notag \\
\sigma_{j_2}(V_2) &=& \frac{i e^2 T_2^2}{2\pi \hbar^2} \sum_{\sigma_2 }
G_J(j_2\sigma_2, -\frac{1}{\hbar}e V_2) A_{d_2} (0) . \label{eqn3.2.5}
\end{eqnarray}
In the Fermi liquid state, these expressions reduce to (via Eq. (\ref{eqn2.3.4}))
\begin{eqnarray}
\sigma_{j_1}(V_1) &=& \frac{4 \pi e^2 T_1^2}{\hbar} \rho_c(-e V_1) \rho_{d_1} (0) , \notag \\
\sigma_{j_2}(V_2) &=& \frac{4 \pi e^2 T_2^2}{\hbar} \rho_c(-e V_2) \rho_{d_2} (0) . \label{eqn3.2.6}
\end{eqnarray}

Now, let us calculate the exchange contributions $G^{(2)}_{J,E}$ and $\sigma^{(E)}_{j_1 j_2}$. The exchange term $G^{(2)}_{J,E}$ can be partitioned into four components
\begin{equation}
G^{(2)}_{J,E} = \sum_{a_1 a_2 =\pm} G^{(a_1 a_2)}_{J,E} , \label{eqn3.2.7}
\end{equation}
where, following Eq. (\ref{eqn3.1.48}), $G^{(a_1 a_2)}_{J,E}$ are defined by 
\begin{eqnarray}
&&G^{(a_1 a_2)}_{J,E}(j_1\sigma_1,j_2\sigma_2; \omega_1, \omega_2)  \notag \\
&=& \iint_c d t_2 d t_1
\wick{ \langle T_c J^{(a_1)}_{1A} (j_1\sigma_1; \c1 t, \c2 t_1; \omega_1) J^{(a_2)}_{2A} (j_2\sigma_2; \c2 t, \c1 t_2; \omega_2) \rangle }. \notag \\
&& \label{eqn3.2.8}
\end{eqnarray}

Applying Wick's theorem \citep{Langreth1976,Rammer2007,HaugJauho2008}, the exchange component $G^{(++)}_{J,E}$ can be evaluated as 
\begin{eqnarray}
&&G^{(+ +)}_{J,E}(j_1\sigma_1,j_2\sigma_2; \omega_1, \omega_2)  \notag \\
&=& \iint_c d t_2 d t_1 \wick{ \langle T_c J^{(+)}_{1A} (j_1\sigma_1; \c1 t, \c2 t_1; \omega_1) J^{(+)}_{2A} (j_2\sigma_2; \c2 t, \c1 t_2; \omega_2) \rangle } \notag \\
&=& (-i)^2 \iint_c d t_2 d t_1 
\wick{
 \langle T_c \c1 c_{j_1\sigma_1}(t) \c2 c^\dag_{j_1\sigma_1}(t_1) \c2 c_{j_2\sigma_2}(t) \c1 c^\dag_{j_2\sigma_2}(t_2) \rangle
} \notag \\
&& \times e^{-i\omega_1 (t_1 - t) - i\omega_2 (t_2 -t)} \notag \\
&=& - \iint_c d t_2 d t_1 G_c(j_1\sigma_1 t; j_2\sigma_2 t_2) \cdot  G_c(j_2\sigma_2 t; j_1\sigma_1 t_1) \notag \\
&& \times e^{-i\omega_1 (t_1 - t) - i\omega_2 (t_2 -t)} . \label{eqn3.2.9}
\end{eqnarray}
Here, the contour-ordered Green's function is defined as 
\begin{equation}
G_c(i\sigma_1 t_1; j\sigma_2 t_2) = -i \langle T_c c_{i\sigma_1}(t_1) c^\dag_{j\sigma_2}(t_2) \rangle . \label{eqn3.2.10}
\end{equation} 
Physically, $G_c(i\sigma_1 t_1; j\sigma_2 t_2)$ describes the quantum propagation amplitude for an electron with spin $\sigma_2$ at location $j_2$ and time $t_2$ to propagate to location $j_1$ with spin $\sigma_1$ at time $t_1$.  Consequently, the exchange term $G^{(++)}_{J,E}$ incorporates two distinct electron propagation processes: (i) The propagation of a spin-$\sigma_1$ electron from $(j_1, t_1)$ to a spin-$\sigma_2$ state at $(j_2, t)$, governed by the $\omega_1$ dynamics, and (ii) the propagation of a spin-$\sigma_2$ electron from $(j_2, t_2)$ to a spin-$\sigma_1$ state at $(j_1, t)$, governed by $\omega_2$ dynamics.
In the Fermi liquid state, the contour-ordered Green’s function takes the form
\begin{eqnarray}
&& G_c(i\sigma_1 t_1; j\sigma_2 t_2) \notag \\
&=& \delta_{\sigma_1 \sigma_2}\frac{1}{N}\sum_{\mathbf{k}} \{ -i\theta_c(t_1-t_2)[1-n_F(\varepsilon_{\mathbf{k}})] + i\theta_c(t_2-t_1) \notag \\
&& \times n_F(\varepsilon_{\mathbf{k}}) \} \, e^{i\mathbf{k}\cdot (\mathbf{R}_i-\mathbf{R}_j)}\cdot e^{i\varepsilon_{\mathbf{k}}(t_2-t_1)/\hbar} . \label{eqn3.2.11}
\end{eqnarray}
For the special case $t_1=t_2=t$, we adopt the limit  $t_2=t_c^+$, where $t_c^+$  is infinitesimally later than $t$  along the time contour. This yields
\begin{equation}
G_c(i\sigma_1 t; j\sigma_2 t_c^+) = \delta_{\sigma_1 \sigma_2}\frac{1}{N}\sum_{\mathbf{k}} [i \, n_F(\varepsilon_{\mathbf{k}})] e^{i\mathbf{k}\cdot (\mathbf{R}_i-\mathbf{R}_j)} . \label{eqn3.2.12}
\end{equation}
Let us define the Green's function
\begin{equation}
\mathscr{G}_R(\omega, \Delta \mathbf{R}) = \frac{1}{N} \sum_{\mathbf{k}} \frac{e^{i\mathbf{k}\cdot \Delta \mathbf{R}} }{\omega -\varepsilon_{\mathbf{k}}/\hbar + i \delta^+ } , \label{eqn3.2.13}
\end{equation}
where $\Delta \mathbf{R}=\mathbf{R}_{j_1}-\mathbf{R}_{j_2}$ denotes the spatial separation. This function characterizes the electron propagation from $\mathbf{R}_{j_2}$ to $\mathbf{R}_{j_1}$ with energy $\hbar \omega$. Using these results, the exchange term $G^{(+ +)}_{J,E}$ simplifies to
\begin{eqnarray}
&&G^{(+ +)}_{J,E}(j_1\sigma_1,j_2\sigma_2; \omega_1, \omega_2)  \notag \\
&=& -\delta_{\sigma_1 \sigma_2} \mathscr{G}_R(\omega_1, -\Delta \mathbf{R}) \mathscr{G}_R(\omega_2, \Delta \mathbf{R}) .\label{eqn3.2.14}
\end{eqnarray}
Thus, $G^{(+ +)}_{J,E}$ incorporates two distinct electron propagation processes: (i) from $\mathbf{R}_{j_1}$ to $\mathbf{R}_{j_2}$ under $\omega_1$ dynamics, and (ii) from $\mathbf{R}_{j_2}$ to $\mathbf{R}_{j_1}$ under $\omega_2$ dynamics. 

Similarly, the exchange component $G^{(- -)}_{J,E}$ satisfies 
\begin{equation}
G^{(- -)}_{J,E}(j_1\sigma_1,j_2\sigma_2; \omega_1, \omega_2) = [G^{(+ +)}_{J,E}(j_1\sigma_1,j_2\sigma_2; \omega_1, \omega_2)]^{*} , \label{eqn3.2.15}
\end{equation}
which explicitly evaluates to
\begin{eqnarray}
&&G^{(--)}_{J,E}(j_1\sigma_1,j_2\sigma_2; \omega_1, \omega_2)  \notag \\
&=& -\delta_{\sigma_1 \sigma_2} [\mathscr{G}_R(\omega_1, -\Delta \mathbf{R}) \mathscr{G}_R(\omega_2, \Delta \mathbf{R})]^{*} .\label{eqn3.2.16}
\end{eqnarray}
Physically, $G^{(--)}_{J,E}$ describes the time-reversed counterparts of the propagation processes in $G^{(++)}_{J,E}$: (i) from $\mathbf{R}_{j_2}$ to $\mathbf{R}_{j_1}$ under $\omega_1$ dynamics, and (ii) from $\mathbf{R}_{j_1}$ to $\mathbf{R}_{j_2}$ under $\omega_2$ dynamics. 

From its definition, $G^{(+ -)}_{J,E}$ can be calculated as 
\begin{eqnarray}
&&G^{(+ -)}_{J,E}(j_1\sigma_1,j_2\sigma_2; \omega_1, \omega_2)  \notag \\
&=& \iint_c d t_2 d t_1 \wick{ \langle T_c J^{(+)}_{1A} (j_1\sigma_1; \c1 t, \c2 t_1; \omega_1) J^{(-)}_{2A} (j_2\sigma_2; \c2 t, \c1 t_2; \omega_2) \rangle  } \notag \\
&=& \iint_c d t_2 d t_1 
\wick{
 \langle T_c \c1 c_{j_1\sigma_1}(t) \c2 c^\dag_{j_1\sigma_1}(t_1) \c2 c_{j_2\sigma_2}(t_2) \c1 c^\dag_{j_2\sigma_2}(t) \rangle
} \notag \\
&& \times e^{-i\omega_1 (t_1 - t) + i\omega_2 (t_2 -t)} \notag \\
&=& \iint_c d t_2 d t_1 G_c(j_1\sigma_1 t; j_2\sigma_2 t^+_c) \cdot  G_c(j_2\sigma_2 t_2; j_1\sigma_1 t_1) \notag \\
&& \times e^{-i\omega_1 (t_1 - t) + i\omega_2 (t_2 -t)} . \label{eqn3.2.17}
\end{eqnarray}
With a lengthy and similar derivation, we show that $G^{(+ -)}_{J,E}$ vanishes 
\begin{equation}
G^{(+ -)}_{J,E}(j_1\sigma_1,j_2\sigma_2; \omega_1, \omega_2) = 0 . \label{eqn3.2.18}
\end{equation}
Analogously, we show that $G^{(- +)}_{J,E}$ also vanishes 
\begin{equation}
G^{(- +)}_{J,E}(j_1\sigma_1,j_2\sigma_2; \omega_1, \omega_2) = 0 . \label{eqn3.2.19}
\end{equation}

Combining the above results for $G^{(2)}_{J,E}$ with the formula for the coincidence dynamical conductance in Eq. (\ref{eqn3.1.50}), the exchange contribution $\sigma^{(E)}_{j_1 j_2}$ can be expressed compactly as
\begin{eqnarray}
\sigma^{(E)}_{j_1 j_2}(V_1, V_2) &=& \left(\frac{e^2 T_1 T_2}{\pi \hbar^2} \right)^2 \operatorname{Re} \left[ \mathscr{G}_R(\overline{\omega}_1, -\Delta \mathbf{R}) \mathscr{G}_R(\overline{\omega}_2, \Delta \mathbf{R}) \right] \notag \\
&& \times A_{d_1}(0) A_{d_2}(0) , \label{eqn3.2.20}
\end{eqnarray}
where $\overline{\omega}_1$ and $\overline{\omega}_2$ are defined in Eq. (\ref{eqn3.1.44}). 
While the direct conductance $\sigma^{(D)}_{j_1 j_2}$ arises from two independent local electron tunneling events at each tip location, the exchange conductance $\sigma^{(E)}_{j_1 j_2}$ emerges from two correlated nonlocal dynamical processes: (i) tunneling electrons detected at tip-1 have propagated nonlocally from tip-2's location, while (ii) those detected at tip-2 have conversely propagated from tip-1's location. 
More precisely, the definition of $\mathscr{G}_R$ reveals that the exchange conductance $\sigma^{(E)}_{j_1 j_2}$ captures two correlated dynamical electron propagation processes: (i) from $\mathbf{R}_{j_1}$ to $\mathbf{R}_{j_2}$ (or vice versa) under $\overline{\omega}_1$ dynamics driven by the bias voltage $V_1$, and (ii) from $\mathbf{R}_{j_2}$ to $\mathbf{R}_{j_1}$ (or vice versa) under $\overline{\omega}_2$ dynamics driven by the voltage $V_2$. Therefore, the exchange conductance $\sigma^{(E)}_{j_1 j_2}$ directly probes two-electron dynamical propagation in the sample, providing direct access to the dynamical two-body correlations that transcend the single-particle physics.

\subsection{Superconducting state} \label{sec3.3}

Let us now consider the double-tip STS of the sample electrons in the superconducting state described by the Hamiltonian in Eq. (\ref{eqn2.4.1}). The total coincidence dynamical conductance also consists of two contributions: the direct term $\sigma_{j_1 j_2}^{(D)}$ and the exchange term $\sigma_{j_1 j_2}^{(E)}$,
\begin{equation}
\sigma_{j_1 j_2}(V_1, V_2) = \sigma^{(D)}_{j_1 j_2}(V_1, V_2) + \sigma^{(E)}_{j_1 j_2}(V_1, V_2) . \label{eqn3.3.1}
\end{equation}
The direct term $\sigma^{(D)}_{j_1 j_2}$ is the product of the individual conductances at two tips as
\begin{equation}
\sigma^{(D)}_{j_1 j_2}(V_1, V_2) = \sigma_{j_1}(V_1) \cdot \sigma_{j_2}(V_2) , \label{eqn3.3.2}
\end{equation}
where the two individual conductances are given by 
\begin{eqnarray}
\sigma_{j_1}(V_1) &=& \frac{4\pi e^2 T_1^2}{N\hbar}\sum_{\mathbf{k}} [\mu_{\mathbf{k}}^2 \delta(e V_1 + E_{\mathbf{k}}) + |\nu_{\mathbf{k}}|^2 \delta(e V_1 - E_{\mathbf{k}}) ] \notag \\
&& \times \rho_{d_1}(0) , \label{eqn3.3.3}  \\
\sigma_{j_2}(V_2) &=& \frac{4\pi e^2 T_2^2}{N\hbar}\sum_{\mathbf{k}} [\mu_{\mathbf{k}}^2 \delta(e V_2 + E_{\mathbf{k}}) + |\nu_{\mathbf{k}}|^2 \delta(e V_2 - E_{\mathbf{k}}) ] \notag \\
&&\times \rho_{d_2}(0) . \label{eqn3.3.4}
\end{eqnarray}

To compute the exchange contributions, we introduce the Nambu spinors 
\begin{equation}
\Phi_{j_1}(t) = 
\left(\begin{array}{l}
\overline{c}_{j_1\uparrow}(t) \\
\overline{c}_{j_1\downarrow}^\dag(t) 
\end{array} \right), \ 
\Phi_{j_1}^\dag (t) = 
\left(\begin{array}{ll}
\overline{c}^\dag_{j_1\uparrow}(t) &   \overline{c}_{j_1\downarrow}(t) 
\end{array} \right) , \label{eqn3.3.5}
\end{equation}
where the modified operators $\overline{c}_{j_1\sigma}(t)$ and $\overline{c}^\dag_{j_1\sigma}(t)$ are defined as 
\begin{equation} 
\overline{c}_{j_1\sigma}(t) = c_{j_1\sigma}(t) e^{i\overline{\omega}_1 t} , \ \overline{c}^\dag_{j_1\sigma}(t) = c^\dag_{j_1\sigma}(t) e^{-i\overline{\omega}_1 t} . \label{eqn3.3.6}
\end{equation}
Here $\overline{\omega}_1$  is given by Eq. (\ref{eqn3.1.44}). We define the current operator $\overline{J}_{1A}$ as
\begin{equation}
\overline{J}_{1A}(j_1; t, t_1; \overline{\omega}_1) = \overline{J}^{(+)}_{1A}(j_1; t, t_1; \overline{\omega}_1) + \overline{J}^{(-)}_{1A}(j_1; t, t_1; \overline{\omega}_1), \label{eqn3.3.7}
\end{equation}
with the components $\overline{J}^{(+)}_{1A}$ and $\overline{J}^{(-)}_{1A}$ are defined by 
\begin{eqnarray}
&&\overline{J}^{(+)}_{1A}(j_1; t, t_1; \overline{\omega}_1) = i \Phi^\dag_{j_1} (t_1) \Phi_{j_1} (t) , \notag \\
&&\overline{J}^{(-)}_{1A}(i; t, t_1; \overline{\omega}_1) = -i \Phi^\dag_{j_1} (t) \Phi_{j_1} (t_1) . \label{eqn3.3.8}
\end{eqnarray}
This operator satisfies 
\begin{equation}
\overline{J}_{1A}(j_1; t, t_1; \overline{\omega}_1) = \sum_{\sigma_1} J_{1A}(j_1\sigma_1; t, t_1; \overline{\omega}_1). \label{eqn3.3.9}
\end{equation}
Analogously, we define the Nambu spinors $\Phi_{j_2} (t)$ and $\Phi_{j_2}^\dag(t)$ by using $\overline{c}_{j_2\sigma}(t)$ and $\overline{c}^\dag_{j_2\sigma}(t)$, the latter of which are defined similarly by Eq. (\ref{eqn3.3.5}) where $\overline{\omega}_1$ is substituted by $\overline{\omega}_2$. Here $\overline{\omega}_2$ has also been defined in Eq. (\ref{eqn3.1.44}).  This yields 
\begin{equation}
\overline{J}_{2A}(j_2; t, t_2; \overline{\omega}_2) = \sum_{\sigma_2} J_{2A}(j_2\sigma_2; t, t_2; \overline{\omega}_2), \label{eqn3.3.10}
\end{equation}
where the current operator $\overline{J}_{2A}$ is defined as
\begin{equation}
\overline{J}_{2A}(j_2; t, t_2; \overline{\omega}_2) = \overline{J}^{(+)}_{2A}(j_2; t, t_2; \overline{\omega}_2) + \overline{J}^{(-)}_{2A}(j_2; t, t_2; \overline{\omega}_2), \label{eqn3.3.11}
\end{equation}
with the components $\overline{J}^{(+)}_{2A}$ and $\overline{J}^{(-)}_{2A}$ given by  
\begin{eqnarray}
&&\overline{J}^{(+)}_{2A}(j_2; t, t_2; \overline{\omega}_2) = i \Phi^\dag_{j_2} (t_2) \Phi_{j_2} (t) , \notag \\
&&\overline{J}^{(-)}_{2A}(j_2; t, t_2; \overline{\omega}_2) = -i \Phi^\dag_{j_2} (t) \Phi_{j_2} (t_2) . \label{eqn3.3.12}
\end{eqnarray}

Using these Nambu spinor representations, the coincidence dynamical conductance becomes
\begin{eqnarray}
&& \sigma_{j_1 j_2}(V_1, V_2) \notag \\
&=& -\left(\frac{e^2 T_1 T_2}{2\pi \hbar^2}\right)^2  \overline{G}^{(2)}_{J}(j_1,j_2; \overline{\omega}_1, \overline{\omega}_2 ) A_{d_1}(0) A_{d_2}(0), \qquad \label{eqn3.3.13}
\end{eqnarray}
where $\overline{\omega}_1$ and $\overline{\omega}_2$ are defined in Eq. (\ref{eqn3.1.44}), and the current correlation function $\overline{G}^{(2)}_{J}$ is defined as
\begin{eqnarray}
&&\overline{G}^{(2)}_{J}(j_1,j_2; \overline{\omega}_1, \overline{\omega}_2)  \notag \\
&=& \iint_c d t_2 d t_1 \langle T_c \overline{J}_{1A} (j_1; t, t_1; \overline{\omega}_1) \overline{J}_{2A} (j_2; t, t_2; \overline{\omega}_2) \rangle . \qquad  \label{eqn3.3.14}
\end{eqnarray}

The exchange contributions to the current correlation function $\overline{G}^{(2)}_{J}$ can be calculated as following: 
\begin{equation}
\overline{G}^{(2)}_{J,E}(j_1,j_2; \overline{\omega}_1, \overline{\omega}_2) = \sum_{a_1 a_2 = \pm} \overline{G}^{(a_1 a_2)}_{J,E}(j_1,j_2; \overline{\omega}_1, \overline{\omega}_2) , \label{eqn3.3.15}
\end{equation}
where the components $\overline{G}^{(a_1 a_2)}_{J,E}$ are defined by
\begin{eqnarray}
&&\overline{G}^{(a_1 a_2)}_{J,E}(j_1,j_2; \overline{\omega}_1, \overline{\omega}_2)  \notag \\
&=& \iint_c d t_2 d t_1
\wick{ \langle T_c \overline{J}^{(a_1)}_{1A} (j_1; \c1 t, \c2 t_1; \overline{\omega}_1) \overline{J}^{(a_2)}_{2A} (j_2; \c2 t, \c1 t_2; \overline{\omega}_2) \rangle }. \, \qquad \label{eqn3.3.16}
\end{eqnarray} 
Explicitly, the component $\overline{G}^{(++)}_{J,E}$ is evaluated by 
\begin{eqnarray}
&&\overline{G}^{(++)}_{J,E}(j_1,j_2; \overline{\omega}_1, \overline{\omega}_2)  \notag \\
&=& \iint_c d t_2 d t_1
\wick{ \langle T_c \overline{J}^{(+)}_{1A} (j_1; \c1 t, \c2 t_1; \overline{\omega}_1) \overline{J}^{(+)}_{2A} (j_2; \c2 t, \c1 t_2; \overline{\omega}_2) \rangle } \notag \\
&=& (i)^2 \iint_c d t_2 d t_1 \wick{ \langle T_c \Phi_{j_1}^\dag \c1 (t_1) \Phi_{j_1} \c2 (t)  \Phi_{j_2}^\dag \c2 (t_2) \Phi_{j_2} \c1 (t) \rangle } \notag \\
&=& -\iint_c d t_2 d t_1 \operatorname{Tr} [\mathcal{G}(j_2 t; j_1 t_1) \mathcal{G}(j_1 t; j_2 t_2)] . \label{eqn3.3.17}
\end{eqnarray}
Here, the Green's functions for the Nambu spinors are defined as
\begin{eqnarray}
&& \mathcal{G}(j_1 t_1; j_2 t_2) = - i \langle T_c \Phi_{j_1}(t_1) \Phi_{j_2}^\dag(t_2) \rangle , \notag \\
&& \mathcal{G}(j_2 t_2; j_1 t_1) = - i \langle T_c \Phi_{j_2}(t_2) \Phi_{j_1}^\dag(t_1) \rangle . \label{eqn3.3.18} 
\end{eqnarray}
Analogously, the other three components are evaluated by 
\begin{eqnarray}
&&\overline{G}^{(--)}_{J,E}(j_1,j_2; \overline{\omega}_1, \overline{\omega}_2)  \notag \\
&=& \iint_c d t_2 d t_1
\wick{ \langle T_c \overline{J}^{(-)}_{1A} (j_1; \c1 t, \c2 t_1; \overline{\omega}_1) \overline{J}^{(-)}_{2A} (j_2; \c2 t, \c1 t_2; \overline{\omega}_2) \rangle } \notag \\
&=& (-i)^2 \iint_c d t_2 d t_1 \wick{ \langle T_c \Phi_{j_1}^\dag \c1 (t) \Phi_{j_1} \c2 (t_1)  \Phi_{j_2}^\dag \c2 (t) \Phi_{j_2} \c1 (t_2) \rangle } \notag \\
&=& -\iint_c d t_2 d t_1 \operatorname{Tr} [\mathcal{G}(j_2 t_2; j_1 t) \mathcal{G}(j_1 t_1; j_2 t)] , \label{eqn3.3.19}
\end{eqnarray}
\begin{eqnarray}
&&\overline{G}^{(+-)}_{J,E}(j_1,j_2; \overline{\omega}_1, \overline{\omega}_2)  \notag \\
&=& \iint_c d t_2 d t_1
\wick{ \langle T_c \overline{J}^{(+)}_{1A} (j_1; \c1 t, \c2 t_1; \overline{\omega}_1) \overline{J}^{(-)}_{2A} (j_2; \c2 t, \c1 t_2; \overline{\omega}_2) \rangle } \notag \\
&=& \iint_c d t_2 d t_1 \wick{ \langle T_c \Phi_{j_1}^\dag \c1 (t_1) \Phi_{j_1} \c2 (t)  \Phi_{j_2}^\dag \c2 (t) \Phi_{j_2} \c1 (t_2) \rangle } \notag \\
&=& -\iint_c d t_2 d t_1 \operatorname{Tr} [\mathcal{G}(j_2 t_2; j_1 t_1) \mathcal{G}(j_1 t; j_2 t)] , \label{eqn3.3.20} 
\end{eqnarray}
\begin{eqnarray}
&&\overline{G}^{(-+)}_{J,E}(j_1,j_2; \overline{\omega}_1, \overline{\omega}_2)  \notag \\
&=& \iint_c d t_2 d t_1
\wick{ \langle T_c \overline{J}^{(-)}_{1A} (j_1; \c1 t, \c2 t_1; \overline{\omega}_1) \overline{J}^{(+)}_{2A} (j_2; \c2 t, \c1 t_2; \overline{\omega}_2) \rangle } \notag \\
&=& \iint_c d t_2 d t_1 \wick{ \langle T_c \Phi_{j_1}^\dag \c1 (t) \Phi_{j_1} \c2 (t_1)  \Phi_{j_2}^\dag \c2 (t_2) \Phi_{j_2} \c1 (t) \rangle } \notag \\
&=& -\iint_c d t_2 d t_1 \operatorname{Tr} [\mathcal{G}(j_2 t; j_1 t) \mathcal{G}(j_1 t_1; j_2 t_2)] . \label{eqn3.3.21}
\end{eqnarray}

From a lengthy and standard derivation, we obtain the current correlation contributions $\overline{G}^{(a_1 a_2)}_{J,E}$ as shown in Appendix \ref{seca2}. We introduce the auxiliary functions
\begin{eqnarray}
&& \mathscr{F}_{11}(\omega, \mathbf{k}) = \frac{\mu_{\mathbf{k}}^2}{\omega-E_{\mathbf{k}}/\hbar+i\delta^+} + \frac{|\nu_{\mathbf{k}}|^2}{\omega+E_{\mathbf{k}}/\hbar+i\delta^+} , \notag \\
&& \mathscr{F}_{12}(\omega, \mathbf{k}) = \frac{\mu_{\mathbf{k}}\nu_{\mathbf{k}}^{*}}{\omega-E_{\mathbf{k}}/\hbar+i\delta^+} - \frac{\mu_{\mathbf{k}}\nu_{\mathbf{k}}^{*}}{\omega+E_{\mathbf{k}}/\hbar+i\delta^+} , \quad \qquad \label{eqn3.3.22}
\end{eqnarray} 
and 
\begin{eqnarray}
&& \mathscr{G}_{11}(\omega; \Delta \mathbf{R}) = \frac{1}{N}\sum_{\mathbf{k}} \mathscr{F}_{11}(\omega, \mathbf{k}) e^{i\mathbf{k}\cdot \Delta \mathbf{R}} , \notag \\
&& \mathscr{G}_{12}(\omega; \Delta \mathbf{R}) = \frac{1}{N}\sum_{\mathbf{k}} \mathscr{F}_{12}(\omega, \mathbf{k}) e^{i\mathbf{k}\cdot \Delta \mathbf{R}} . \label{eqn3.3.23}
\end{eqnarray}
The exchange term $\overline{G}^{(++)}_{J,E}$ evaluates to
\begin{eqnarray}
\overline{G}^{(++)}_{J,E} &=& -[ \mathscr{G}_{11}(\overline{\omega}_1; -\Delta \mathbf{R})\mathscr{G}_{11}(\overline{\omega}_2; \Delta \mathbf{R}) \notag \\
&& + \mathscr{G}_{12}(\overline{\omega}_1; -\Delta \mathbf{R}) \mathscr{G}_{12}^{*}(\overline{\omega}_2; -\Delta \mathbf{R})   \notag \\
&& + \mathscr{G}_{12}^{*}(\overline{\omega}_1; \Delta \mathbf{R}) \mathscr{G}_{12}(\overline{\omega}_2; \Delta \mathbf{R}) \notag \\
&& + \mathscr{G}^{*}_{11}(\overline{\omega}_1; \Delta \mathbf{R}) \mathscr{G}^{*}_{11}(\overline{\omega_2}; -\Delta \mathbf{R}) ] . \label{eqn3.3.24}
\end{eqnarray}
Under spatial inversion symmetry, this simplifies to 
\begin{eqnarray}
\overline{G}^{(++)}_{J,E} &=& -[ \mathscr{G}_{11}(\overline{\omega}_1; -\Delta \mathbf{R})\mathscr{G}_{11}(\overline{\omega}_2; \Delta \mathbf{R}) \notag \\
&& + \mathscr{G}_{12}(\overline{\omega}_1; -\Delta \mathbf{R}) \mathscr{G}_{12}^{*}(\overline{\omega}_2; -\Delta \mathbf{R})  + c.c. ] . \quad \label{eqn3.3.25}
\end{eqnarray}
From the results in Appendix \ref{seca2}, the remaining terms satisfy 
\begin{equation}
\overline{G}^{(--)}_{J,E} = [\overline{G}^{(++)}_{J,E}]^{*}, \ \overline{G}^{(-+)}_{J,E} = [\overline{G}^{(+-)}_{J,E}]^{*}=0  . \label{eqn3.3.26}
\end{equation}

The exchange coincidence dynamical conductance is explicitly given by
\begin{eqnarray}
\sigma^{(E)}_{j_1 j_2}(V_1, V_2) &=& -\frac{1}{2}\left(\frac{e^2 T_1 T_2}{\pi \hbar^2}\right)^2 \operatorname{Re} \left[\overline{G}^{(++)}_{J,E} \right] A_{d_1}(0) A_{d_2}(0) ,  \notag \\
&&  \label{eqn3.3.27}
\end{eqnarray}
where $\overline{G}^{(++)}_{J,E}$ is defined in Eq. (\ref{eqn3.3.24}). The expression for $\overline{G}^{(++)}_{J,E}$ reveals that the exchange conductance $\sigma^{(E)}_{j_1 j_2}$ arises from two distinct channel contributions: (i) the normal-electron terms governed by  $\mathscr{G}_{11}$ and (ii) the superconducting-condensate terms determined by $\mathscr{G}_{12}$. Notably, the latter contribution vanishes in the normal state due to its proportionality to the superconducting gap function. In this case, the exchange conductance $\sigma^{(E)}_{j_1 j_2}$ reverts to the Fermi liquid form given by Eq. (\ref{eqn3.2.20}). Thus, the exchange conductance $\sigma^{(E)}_{j_1 j_2}$ can serve as a probe of nonlocal dynamical two-body correlations of both the normal electrons and the superconducting condensate.

\section{An extended double-tip scanning tunneling spectroscopy} \label{sec4}

\subsection{Basic principle} \label{sec4.1}

In Sec. \ref{sec3}, we introduced a coincidence double-tip STS for measuring two tunneling currents simultaneously. Here, we extend this concept to measurements at different times by defining the observable 
\begin{equation}
\overline{\langle I_1(t_1) I_2(t_2)\rangle}_c = \langle [I_{1H}(t_1) I_{2H}(t_2)]_c \rangle , \label{eqn4.1.1}
\end{equation}
where the symmetrized current product is given by
\begin{equation}
[I_{1H}(t_1) I_{2H}(t_2)]_c = \frac{1}{2}[I_{1H}(t_1) I_{2H}(t_2) + I_{2H}(t_2) I_{1H}(t_1)] . \label{eqn4.1.2}
\end{equation} 
This quantity is real-valued and satisfies
\begin{equation}
\overline{\langle I_1(t_1) I_2(t_2)\rangle}_c = \operatorname{Re} \overline{\langle I_1(t_1) I_2(t_2)\rangle} , \label{eqn4.1.3}
\end{equation} 
with 
\begin{equation}
\overline{\langle I_1(t_1) I_2(t_2)\rangle} = \langle T_c S_c(t_i, t_i) I_{1I}(t_1) I_{2I}(t_2) \rangle . \label{eqn4.1.4}
\end{equation}
Here the time contour is defined as $t_i \rightarrow t_f \rightarrow t_i$ with $t_f =\operatorname{max}\{t_1, t_2\}$. 

To second order in perturbation theory, we obtain 
\begin{eqnarray}
&& \overline{\langle I_1(t_1) I_2(t_2)\rangle} \notag \\
&=& \frac{1}{2} \left(\frac{-i}{\hbar} \right)^2 \iint_c d t_2^\prime d t_1^\prime \langle T_c V_I(t_2^\prime) V_I(t_1^\prime) I_{1I}(t_1) I_{2I}(t_2) \rangle .  \notag \\
&&  \label{eqn4.1.5}
\end{eqnarray}
The correlation function decomposes into four components
\begin{equation}
\overline{\langle I_1(t_1) I_2(t_2)\rangle} = \sum_{a_1 a_2 =\pm} \overline{\langle I_1(t_1) I_2(t_2)\rangle}^{(a_1 a_2)} , \label{eqn4.1.6}
\end{equation}
where each term takes the form
\begin{eqnarray}
&&\overline{\langle I_1(t_1) I_2(t_2)\rangle}^{(a_1 a_2)}  \notag \\
&=& \frac{(e T_1 T_2)^2}{\hbar^4} \iint_c d t_2^\prime d t_1^\prime e^{i a_1 [\phi_1(t_1)-\phi_1(t_1^\prime)]+i a_2[\phi_2(t_2)-\phi_2(t_2^\prime)]} \notag \\
&& \times a_1 a_2 \cdot \langle T_c A_1^{(a_1)}(t_1) A_2^{(a_2)}(t_2) A_2^{(\overline{a}_2)}(t_2^\prime) A_1^{(\overline{a}_1)}(t_1^\prime) \rangle . \,\qquad \label{eqn4.1.7}
\end{eqnarray}
Following the methodology of Sec. \ref{sec3}, we evaluate these terms with the results presented in Appendix \ref{seca3}. 

Introducing the correlation function
\begin{eqnarray}
&& G_{c,l_1 l_2} (j_1\sigma_1 t_1, j_2\sigma_2 t_2; \omega_1, \omega_2) \notag \\
&=& -\sum_{a_1 a_2 =\pm} (-1)^{a_1\cdot a_2} G_{c,l_1 l_2}^{(a_1 a_2)} (j_1\sigma_1 t_1, j_2\sigma_2 t_2; \omega_1, \omega_2) , \,\qquad \label{eqn4.1.8}
\end{eqnarray}
where $G_{c,l_1 l_2}^{(a_1 a_2)} $ are defined in Appendix \ref{seca3}, we derive the current correlation as
\begin{widetext}
\begin{eqnarray}
\overline{\langle I_1(t_1) I_2(t_2)\rangle}_c &=& \left(\frac{e T_1 T_2}{2\pi \hbar^2}\right)^2 \sum_{\sigma_1 \sigma_2 l_1 l_2} \iint d\xi_1 d\xi_2 \operatorname{Re} [ G_{c,l_1 l_2}(j_1\sigma_1 t_1, j_2\sigma_2 t_2; \frac{1}{\hbar}(\xi_1 - eV_1), \frac{1}{\hbar}(\xi_2 - eV_2))] \notag \\
&& \times G^{(l_1)}_{d_1}(\xi_1) G^{(l_2)}_{d_2}(\xi_2) A_{d_1}(\xi_1) A_{d_2}(\xi_2) . \label{eqn4.1.9}
\end{eqnarray}
\end{widetext}
This expression parallels Eq. (\ref{eqn3.1.39}) for simultaneous coincidence double-tip STS.

We define the coincidence dynamical conductance for the extended double-tip STS as  
\begin{equation}
\sigma_{j_1 t_1,j_2 t_2}(V_1, V_2) = \frac{d}{d V_2} \frac{d}{d V_1} \overline{\langle I_1(t_1) I_2(t_2)\rangle}_c . \label{eqn4.1.10}
\end{equation}
It evaluates to 
\begin{eqnarray}
&& \sigma_{j_1 t_1,j_2 t_2}(V_1, V_2) \notag \\
&=& -\left(\frac{e^2 T_1 T_2}{2\pi \hbar^2}\right)^2 \sum_{\sigma_1 \sigma_2} \operatorname{Re} [ G^{(2)}_{J}(j_1\sigma_1 t_1, j_2\sigma_2 t_2; \overline{\omega}_1, \overline{\omega}_2 )] \notag \\
&& \times A_{d_1}(0) A_{d_2}(0),  \label{eqn4.1.11}
\end{eqnarray}
where $\overline{\omega}_1$ and $\overline{\omega}_2$ are given in Eq. (\ref{eqn3.1.44}), and $G^{(2)}_{J}$ represents the second-order current correlation function
\begin{eqnarray}
&& G^{(2)}_{J}(j_1\sigma_1 t_1,j_2\sigma_2 t_2; \omega_1, \omega_2) \notag \\
&=& \iint_c d t_2^\prime d t_1^\prime \langle T_c J_{1A} (j_1\sigma_1; t_1, t_1^\prime; \omega_1) J_{2A} (j_2\sigma_2; t_2, t_2^\prime; \omega_2) \rangle .  \notag \\
&& \label{eqn4.1.12}
\end{eqnarray}
Notably, when $t_1=t_2$ , Eq. (\ref{eqn4.1.11}) reduces to the simultaneous coincidence measurement result given in Eq. (\ref{eqn3.1.50}), demonstrating consistency between the extended and simultaneous coincidence double-tip STS formalisms.

\subsection{The Fermi liquid state} \label{sec4.2}

Let us study the coincidence dynamical conductance of the extended double-tip STS for the sample electrons in the Fermi liquid state defined by the Hamiltonian in Eq. (\ref{eqn2.3.1}). It also comprises direct and exchange contributions
\begin{equation}
\sigma_{j_1 t_1,j_2 t_2} (V_1, V_2) = \sigma^{(D)}_{j_1 t_1,j_2 t_2} (V_1, V_2) + \sigma^{(E)}_{j_1 t_1,j_2 t_2} (V_1, V_2) . \label{eqn4.2.1}
\end{equation}
The direct term maintains time independence
\begin{equation}
\sigma^{(D)}_{j_1 t_1,j_2 t_2} (V_1, V_2) = \sigma_{j_1} (V_1) \cdot \sigma_{j_2} (V_2) , \label{eqn4.2.2}
\end{equation}
where $\sigma_{j_1} (V_1)$ and $\sigma_{j_2} (V_2)$ are given in Eq. (\ref{eqn3.2.5}).  

The exchange contribution takes the form
\begin{eqnarray}
&& \sigma^{(E)}_{j_1 t_1,j_2 t_2}(V_1, V_2) \notag \\
&=& -\left(\frac{e^2 T_1 T_2}{2\pi \hbar^2}\right)^2 \sum_{\sigma_1 \sigma_2} \operatorname{Re} [ G^{(2)}_{J,E}(j_1\sigma_1 t_1, j_2\sigma_2 t_2; \overline{\omega}_1, \overline{\omega}_2 )] \notag \\
&& \times A_{d_1}(0) A_{d_2}(0),  \label{eqn4.2.3}
\end{eqnarray}
with the exchange correlation function $G^{(2)}_{J,E}$ defined as
\begin{eqnarray}
&& G^{(2)}_{J,E}(j_1\sigma_1 t_1,j_2\sigma_2 t_2; \omega_1, \omega_2) \notag \\
&=& \iint_c d t_2^\prime d t_1^\prime \wick{ \langle T_c J_{1A} (j_1\sigma_1; \c1 t_1, \c2 t_1^\prime; \omega_1) J_{2A} (j_2\sigma_2; \c2 t_2, \c1 t_2^\prime; \omega_2) \rangle} .  \notag \\
&& \label{eqn4.2.4}
\end{eqnarray}
For the Fermi liquid state, we evaluate $G^{(2)}_{J,E}$ in Appendix \ref{seca4}. 
By introducing the time-dependent phase factor
\begin{equation}
\mathscr{T}_1(\Delta t) = e^{i(\overline{\omega}_2 - \overline{\omega}_1) \Delta t}, \quad \Delta t = t_2 - t_1,  \label{eqn4.2.5}
\end{equation}
and employing the function $\mathscr{G}_R(\omega, \Delta \mathbf{R})$ defined in Eq. (\ref{eqn3.2.13}), we obtain the exchange conductance as
\begin{eqnarray}
 && \sigma^{(E)}_{j_1 t_1,j_2 t_2}(V_1, V_2) \notag \\
&=& \left(\frac{e^2 T_1 T_2}{\pi\hbar^2}\right)^2 \operatorname{Re}[\mathscr{T}_1(\Delta t) \mathscr{G}_R(\overline{\omega}_1, -\Delta \mathbf{R}) \mathscr{G}_R(\overline{\omega}_2, \Delta \mathbf{R})] \notag  \\
&& \times  A_{d_1}(0) A_{d_2}(0) . \label{eqn4.2.6}
\end{eqnarray}
This result differs from the simultaneous coincidence measurement case in Eq. (\ref{eqn3.2.20}) through the inclusion of the temporal phase factor $\mathscr{T}_1(\Delta t)$. The extended formulation reduces exactly to the simultaneous coincidence result when $\Delta t=0$ ($t_1=t_2$), thereby providing a consistent generalization of the coincidence double-tip STS formalism to non-simultaneous coincidence measurements.

\section{Discussions and conclusion} \label{sec5}

In this work, we propose a coincidence double-tip STS for spatially resolved measurement of dynamical two-body correlations. This technique extends conventional single-tip STS by using a double-tip STM setup, where two independently biased tips (with voltages $V_1$ and $V_2$) simultaneously measure quantum tunneling currents $I_1(t)$ and $I_2(t)$ at distinct locations $j_1$ and $j_2$. From the coincidence tunneling current correlation $\overline{\langle I_1(t) I_2(t)\rangle}$, we define a coincidence dynamical conductance via differentiation with respect to the bias voltages $V_1$ and $V_2$. Our nonequilibrium theory reveals that the coincidence dynamical conductance directly probes a contour-ordered second-order current correlation function, offering a powerful approach to study strongly correlated electron systems. 

To demonstrate the capabilities of the coincidence double-tip STS, we have applied the nonequilibrium theory to calculate the coincidence dynamical conductances for two electronic states, a nearly free Fermi liquid and a superconducting state. The coincidence dynamical conductance comprises two distinct components, the direct conductance and the exchange conductance. The direct conductance arises as a product of the individual single-tip STS dynamical conductances associated with each probing tip, reflecting independent local tunneling processes. The exchange conductance is a  fundamentally nonlocal contribution which captures correlated electron propagations between the two tips, where (i) electrons detected at tip-1 originate from tip-2’s location, and (ii) conversely, electrons detected at tip-2 propagate from tip-1’s location. Crucially, the exchange conductance can provide direct experimental access to dynamical two-body correlations that cannot be described within single-particle physics. In superconducting systems, this term reveals particularly rich physics as it incorporates both the conventional normal electron propagations and the superconducting two-particle processes arising from the condensates. Thus, through the exchange conductance, the coincidence double-tip STS offers unique insights into the dynamical two-body properties of superconducting condensates.

We further generalize this technique to time-resolved measurement by introducing a temporal delay $\Delta t=t_2 - t_1$ between the two tunneling current detections. The resulting extended coincidence dynamical conductance acquires a phase factor $\exp (i\Delta \overline{\omega} \cdot \Delta t)$ where $\Delta \overline{\omega} = e(V_1-V_2)/\hbar$. This formalism imposes the following experimental condition for simultaneous coincidence double-tip STS measurements 
\begin{equation}
| \Delta \overline{\omega} \cdot \Delta t | \ll 1 . \label{eqn5.1}
\end{equation} 

In most strongly correlated electron systems, the dominant correlations originate from highly localized physics, typically characterized by correlation lengths comparable to interatomic distances (a few $\text{\AA}$). This presents significant experimental challenge for resolving two-body nonlocal correlations using coincidence double-tip STS, as it requires precise control of tip separations at these extremely short length scales. The recent discovery of unconventional superconductivity and Mott physics in magic-angle twisted bilayer graphene \citep{CaoNature2018} offers a promising alternative platform. In this system, the exotic correlated phenomena are intimately relevant to the moir\'{e} superlattice, whose characteristic periodicity ($\sim 13~\text{nm}$) defines a mesoscopic length scale. This length scale is amenable to investigation by coincidence double-tip STS, as modern double-tip STM setups have successfully achieved tip separations within the $\sim 30~\text{nm}$ range \citep{NakayamaAdM201200257, MaartenNature2020}, making such measurements experimentally feasible. Beyond graphene-based moir\'{e} systems, the broader family of twisted van der Waals heterostructures, including twisted transition metal dichalcogenides \citep{TangNature2020, ReganNature2020}, offers a continuously tunable platform in which the moir\'{e} period and the interaction strength can be engineered on demand, thereby expanding the materials base available for coincidence double-tip STS investigations of strongly correlated electron systems. Taken together, moir\'{e} quantum materials emerge as an ideal platform for implementing coincidence double-tip STS to probe nonlocal dynamical two-body correlations in strongly correlated electron systems.

It is worth noting that the experimental realization of coincidence double-tip STS is being actively advanced by ongoing technological progress. Conventional tips are fabricated from wires with diameters of tens to hundreds of micrometers, which significantly prevents the minimum achievable tip separation to the order of tens of nanometers.  This limitation is being overcome by the development of microfabricated probe tips, such as carbon-nanotube-terminated and graphene-based probe tips \citep{NakayamaAdM201200257, MaartenNature2020}, which have already achieved tip separations to a few tens of nanometers. Further advances in nanofabrication are expected to push the accessible tip separations into the nanometer regime. Beyond its immediate experimental implementation, the nonequilibrium theoretical framework developed in this work plays an essential role in guiding future studies of strongly correlated electron systems. By establishing a rigorous connection between the measured coincidence dynamical conductance and the underlying contour-ordered second-order current correlation function, the theory developed here defines precisely which dynamical two-body correlations are accessible and how they can be extracted from experimental data. The nonequilibrium theoretical framework therefore not only motivates the technological development of double-tip STS instrumentation but also supplies the interpretive framework necessary for future experiments to translate raw current-current correlation data into microscopic insight. In this sense, coincidence double-tip STS is poised to become a highly promising tool for unraveling the dynamical two-body physics of strongly correlated electron systems.

In summary, building upon the nonequilibrium theoretical framework developed in this work, we propose a coincidence double-tip STS for direct measurement of spatially resolved dynamical two-body correlations. This technique provides a powerful tool for investigating the exotic physics of strongly correlated electron systems.

\acknowledgments
We thank Prof. Yunan Yan and Prof. Boye Sun for invaluable discussions. 
This work was supported by the National Natural Science Foundation of China (Grants No. 12674172 and No. 11874318) and the Natural Science Foundation of Shandong Province (Grant No. ZR2023MA015).



\appendix 

\begin{widetext}

\section{Calculation of $\overline{\langle I_1(t) I_2(t)\rangle}^{(a_1 a_2)}$ } \label{seca1}

Following analogous methodology to the calculation of $\overline{\langle I_1(t) I_2(t)\rangle}^{(+ +)}$, we derive the remaining three components $\overline{\langle I_1(t) I_2(t)\rangle}^{(a_1 a_2)}$  as shown below.
\begin{eqnarray}
\overline{\langle I_1(t) I_2(t)\rangle}^{(--)} &=& -\left(\frac{e T_1 T_2}{2\pi \hbar^2}\right)^2 \sum_{\sigma_1 \sigma_2 l_1 l_2} \iint d\xi_1 d\xi_2 G^{(--)}_{c,l_1 l_2}(j_1\sigma_1, j_2\sigma_2; \frac{1}{\hbar}(\xi_1 - eV_1), \frac{1}{\hbar}(\xi_2 - eV_2)) G^{(l_1)}_{d_1}(\xi_1) G^{(l_2)}_{d_2}(\xi_2) \notag \\
&& \times A_{d_1}(\xi_1) A_{d_2}(\xi_2) , \label{eqna1.1}
\end{eqnarray}
where 
\begin{equation}
G^{(--)}_{c,l_1 l_2}(j_1\sigma_1, j_2\sigma_2; \omega_1, \omega_2) = \iint_c d t_2 d t_1 \overline{\theta}^{(l_1)}_c(t-t_1) \overline{\theta}^{(l_2)}_c(t-t_2) G_c(j_1\sigma_1 t_1, j_2\sigma_2 t_2; j_2\sigma_2 t, j_1\sigma_1 t) e^{i\omega_1(t_1-t)+i\omega_2 (t_2-t)} . \label{eqna1.2} 
\end{equation}
\begin{eqnarray}
\overline{\langle I_1(t) I_2(t)\rangle}^{(+-)} &=& \left(\frac{e T_1 T_2}{2\pi \hbar^2}\right)^2 \sum_{\sigma_1 \sigma_2 l_1 l_2} \iint d\xi_1 d\xi_2 G^{(+-)}_{c,l_1 l_2}(j_1\sigma_1, j_2\sigma_2; \frac{1}{\hbar}(\xi_1 - eV_1), \frac{1}{\hbar}(\xi_2 - eV_2)) G^{(l_1)}_{d_1}(\xi_1) G^{(l_2)}_{d_2}(\xi_2) \notag \\
&& \times A_{d_1}(\xi_1) A_{d_2}(\xi_2) , \label{eqna1.3}
\end{eqnarray}
where 
\begin{equation}
G^{(+-)}_{c,l_1 l_2}(j_1\sigma_1, j_2\sigma_2; \omega_1, \omega_2) = \iint_c d t_2 d t_1 \overline{\theta}^{(l_1)}_c(t_1-t) \overline{\theta}^{(l_2)}_c(t-t_2) G_c(j_1\sigma_1 t, j_2\sigma_2 t_2; j_2\sigma_2 t, j_1\sigma_1 t_1) e^{-i\omega_1(t_1-t)+i\omega_2 (t_2-t)} . \label{eqna1.4} 
\end{equation}
\begin{eqnarray}
\overline{\langle I_1(t) I_2(t)\rangle}^{(-+)} &=& \left(\frac{e T_1 T_2}{2\pi \hbar^2}\right)^2 \sum_{\sigma_1 \sigma_2 l_1 l_2} \iint d\xi_1 d\xi_2 G^{(-+)}_{c,l_1 l_2}(j_1\sigma_1, j_2\sigma_2; \frac{1}{\hbar}(\xi_1 - eV_1), \frac{1}{\hbar}(\xi_2 - eV_2)) G^{(l_1)}_{d_1}(\xi_1) G^{(l_2)}_{d_2}(\xi_2) \notag \\
&& \times A_{d_1}(\xi_1) A_{d_2}(\xi_2) , \label{eqna1.5}
\end{eqnarray}
where 
\begin{equation}
G^{(-+)}_{c,l_1 l_2}(j_1\sigma_1, j_2\sigma_2; \omega_1, \omega_2) = \iint_c d t_2 d t_1 \overline{\theta}^{(l_1)}_c(t-t_1) \overline{\theta}^{(l_2)}_c(t_2-t) G_c(j_1\sigma_1 t_1, j_2\sigma_2 t; j_2\sigma_2 t_2, j_1\sigma_1 t) e^{i\omega_1(t_1-t)-i\omega_2 (t_2-t)} . \label{eqna1.6} 
\end{equation}

\section{Calculation of $\overline{G}^{(2)}_{J,E}$ in superconducting state} \label{seca2}

We begin by evaluating the Nambu spinor Green's functions introduced in Eq. (\ref{eqn3.3.18}). They can be decomposed into greater and lesser components as 
\begin{equation}
\mathcal{G}(j_1 t_1; j_2 t_2) = \theta_c(t_1-t_2) \mathcal{G}^{>}(j_1 t_1; j_2 t_2) + \theta_c(t_2-t_1) \mathcal{G}^{<}(j_1 t_1; j_2 t_2) , \label{eqna2.1}
\end{equation}
where the greater and lesser Green's functions, $\mathcal{G}^{>}$ and $\mathcal{G}^{<}$, are respectively defined by 
\begin{eqnarray}
&&\mathcal{G}^{>}(j_1 t_1; j_2 t_2) = -i \left(
\begin{array}{ll}
\langle \overline{c}_{j_1\uparrow}(t_1) \overline{c}^\dag_{j_2\uparrow}(t_2) \rangle & \langle \overline{c}_{j_1\uparrow}(t_1) \overline{c}_{j_2\downarrow}(t_2) \rangle \\
\langle \overline{c}^\dag_{j_1\downarrow}(t_1) \overline{c}^\dag_{j_2\uparrow}(t_2) \rangle & \langle \overline{c}^\dag_{j_1\downarrow}(t_1) \overline{c}_{j_2\downarrow}(t_2) \rangle
\end{array} \right) ,   \notag \\
&&\mathcal{G}^{<}(j_1 t_1; j_2 t_2) = i\left(
\begin{array}{ll}
\langle \overline{c}^\dag_{j_2\uparrow}(t_2) \overline{c}_{j_1\uparrow}(t_1) \rangle & \langle \overline{c}_{j_2\downarrow}(t_2) \overline{c}_{j_1\uparrow}(t_1)  \rangle \\
\langle \overline{c}^\dag_{j_2\uparrow}(t_2) \overline{c}^\dag_{j_1\downarrow}(t_1) \rangle & \langle \overline{c}_{j_2\downarrow}(t_2) \overline{c}^\dag_{j_1\downarrow}(t_1) \rangle
\end{array} \right) . \label{eqna2.2}
\end{eqnarray}

Through direct calculation, we demonstrate that
\begin{eqnarray}
&&\mathcal{G}^{>}_{11}(j_1 t_1; j_2 t_2) \notag \\
&=& -\frac{i}{N}\sum_{\mathbf{k}} \{ \mu^2_{\mathbf{k}}[1-n_F(E_\mathbf{k})]e^{i(\overline{\omega}_1-E_{\mathbf{k}}/\hbar )t_1 - i(\overline{\omega}_2 - E_{\mathbf{k}}/\hbar ) t_2} + |\nu_{\mathbf{k}}|^2 n_F(E_\mathbf{k}) e^{i(\overline{\omega}_1+E_{\mathbf{k}}/\hbar )t_1 - i(\overline{\omega}_2 + E_{\mathbf{k}}/\hbar ) t_2}  \} e^{i\mathbf{k}\cdot (\mathbf{R}_{j_1}-\mathbf{R}_{j_2})} , \notag 
\end{eqnarray}
\begin{eqnarray}
&&\mathcal{G}^{>}_{12}(j_1 t_1; j_2 t_2) \notag \\
&=& -\frac{i}{N}\sum_{\mathbf{k}} \mu_{\mathbf{k}}\nu_{\mathbf{k}} \{ [1-n_F(E_\mathbf{k})]e^{i(\overline{\omega}_1-E_{\mathbf{k}}/\hbar )t_1 + i(\overline{\omega}_2 + E_{\mathbf{k}}/\hbar ) t_2} - n_F(E_\mathbf{k}) e^{i(\overline{\omega}_1+E_{\mathbf{k}}/\hbar )t_1 + i(\overline{\omega}_2 - E_{\mathbf{k}}/\hbar ) t_2}  \} e^{i\mathbf{k}\cdot (\mathbf{R}_{j_1}-\mathbf{R}_{j_2})} , \notag 
\end{eqnarray}
\begin{eqnarray}
&&\mathcal{G}^{>}_{21}(j_1 t_1; j_2 t_2) \notag \\
&=& -\frac{i}{N}\sum_{\mathbf{k}} \mu_{\mathbf{k}}\nu^{*}_{\mathbf{k}} \{ [1-n_F(E_\mathbf{k})]e^{-i(\overline{\omega}_1+E_{\mathbf{k}}/\hbar )t_1 - i(\overline{\omega}_2 - E_{\mathbf{k}}/\hbar ) t_2} - n_F(E_\mathbf{k}) e^{-i(\overline{\omega}_1-E_{\mathbf{k}}/\hbar )t_1 - i(\overline{\omega}_2 + E_{\mathbf{k}}/\hbar ) t_2}  \} e^{i\mathbf{k}\cdot (\mathbf{R}_{j_1}-\mathbf{R}_{j_2})} , \notag 
\end{eqnarray}
\begin{eqnarray}
&&\mathcal{G}^{>}_{22}(j_1 t_1; j_2 t_2) \notag \\
&=& -\frac{i}{N}\sum_{\mathbf{k}} \{ |\nu_{\mathbf{k}}|^2 [1-n_F(E_\mathbf{k})]e^{-i(\overline{\omega}_1+E_{\mathbf{k}}/\hbar )t_1 + i(\overline{\omega}_2 + E_{\mathbf{k}}/\hbar ) t_2} + \mu_{\mathbf{k}}^2 n_F(E_\mathbf{k}) e^{-i(\overline{\omega}_1-E_{\mathbf{k}}/\hbar )t_1 + i(\overline{\omega}_2 - E_{\mathbf{k}}/\hbar ) t_2}  \} e^{i\mathbf{k}\cdot (\mathbf{R}_{j_1}-\mathbf{R}_{j_2})} , \notag \\
&& \label{eqna2.3}
\end{eqnarray}
and 
\begin{eqnarray}
&&\mathcal{G}^{<}_{11}(j_1 t_1; j_2 t_2) \notag \\
&=& \frac{i}{N}\sum_{\mathbf{k}} \{ \mu^2_{\mathbf{k}}n_F(E_\mathbf{k})e^{i(\overline{\omega}_1-E_{\mathbf{k}}/\hbar )t_1 - i(\overline{\omega}_2 - E_{\mathbf{k}}/\hbar ) t_2} + |\nu_{\mathbf{k}}|^2 [1- n_F(E_\mathbf{k})] e^{i(\overline{\omega}_1+E_{\mathbf{k}}/\hbar )t_1 - i(\overline{\omega}_2 + E_{\mathbf{k}}/\hbar ) t_2}  \} e^{i\mathbf{k}\cdot (\mathbf{R}_{j_1}-\mathbf{R}_{j_2})} , \notag 
\end{eqnarray}
\begin{eqnarray}
&&\mathcal{G}^{<}_{12}(j_1 t_1; j_2 t_2) \notag \\
&=& \frac{i}{N}\sum_{\mathbf{k}} \mu_{\mathbf{k}}\nu_{\mathbf{k}} \{ n_F(E_\mathbf{k}) e^{i(\overline{\omega}_1-E_{\mathbf{k}}/\hbar )t_1 + i(\overline{\omega}_2 + E_{\mathbf{k}}/\hbar ) t_2} - [1- n_F(E_\mathbf{k})] e^{i(\overline{\omega}_1+E_{\mathbf{k}}/\hbar )t_1 + i(\overline{\omega}_2 - E_{\mathbf{k}}/\hbar ) t_2}  \} e^{i\mathbf{k}\cdot (\mathbf{R}_{j_1}-\mathbf{R}_{j_2})} , \notag 
\end{eqnarray}
\begin{eqnarray}
&&\mathcal{G}^{<}_{21}(j_1 t_1; j_2 t_2) \notag \\
&=& \frac{i}{N}\sum_{\mathbf{k}} \mu_{\mathbf{k}}\nu^{*}_{\mathbf{k}} \{ n_F(E_\mathbf{k}) e^{-i(\overline{\omega}_1+E_{\mathbf{k}}/\hbar )t_1 - i(\overline{\omega}_2 - E_{\mathbf{k}}/\hbar ) t_2} - [1-n_F(E_\mathbf{k})] e^{-i(\overline{\omega}_1-E_{\mathbf{k}}/\hbar )t_1 - i(\overline{\omega}_2 + E_{\mathbf{k}}/\hbar ) t_2}  \} e^{i\mathbf{k}\cdot (\mathbf{R}_{j_1}-\mathbf{R}_{j_2})} , \notag 
\end{eqnarray}
\begin{eqnarray}
&&\mathcal{G}^{<}_{22}(j_1 t_1; j_2 t_2) \notag \\
&=& \frac{i}{N}\sum_{\mathbf{k}} \{ |\nu_{\mathbf{k}}|^2 n_F(E_\mathbf{k}) e^{-i(\overline{\omega}_1+E_{\mathbf{k}}/\hbar )t_1 + i(\overline{\omega}_2 + E_{\mathbf{k}}/\hbar ) t_2} + \mu_{\mathbf{k}}^2 [1-n_F(E_\mathbf{k})] e^{-i(\overline{\omega}_1-E_{\mathbf{k}}/\hbar )t_1 + i(\overline{\omega}_2 - E_{\mathbf{k}}/\hbar ) t_2}  \} e^{i\mathbf{k}\cdot (\mathbf{R}_{j_1}-\mathbf{R}_{j_2})} . \notag \\
 && \label{eqna2.4}
\end{eqnarray}
The Green's function $\mathcal{G}(j_2 t_2; j_1 t_1)$ can be obtained through an analogous expression to $\mathcal{G}(j_1 t_1; j_2 t_2)$ by making the following substitutions: $j_1\leftrightarrow j_2, t_1 \leftrightarrow t_2$ and $\overline{\omega}_1 \leftrightarrow \overline{\omega}_2$. 

For the equal-time case ($t_1 = t_2 =t$), we adopt the following matrix representations:  
\begin{equation}
\mathcal{G}(j_1 t; j_2 t) = \left( 
\begin{array}{ll}
\mathcal{G}^{<}_{11}(j_1 t^{-}_c; j_2 t) & \mathcal{G}^{>}_{12}(j_1 t^{+}_c; j_2 t) \\
\mathcal{G}^{>}_{21}(j_1 t^{+}_c; j_2 t) & \mathcal{G}^{>}_{22}(j_1 t^{+}_c; j_2 t) 
\end{array} \right) , \quad 
\mathcal{G}(j_2 t; j_1 t) = \left( 
\begin{array}{ll}
\mathcal{G}^{<}_{11}(j_2 t^{-}_c; j_1 t) & \mathcal{G}^{>}_{12}(j_2 t^{+}_c; j_1 t) \\
\mathcal{G}^{>}_{21}(j_2 t^{+}_c; j_1 t) & \mathcal{G}^{>}_{22}(j_2 t^{+}_c; j_1 t) 
\end{array} \right) ,  \label{eqna2.5}
\end{equation} 
where $t_c^{+}$ and $t_c^{-}$  denote time points infinitesimally shifted forward and backward, respectively, from $t$ along the time contour.

Through lengthy but standard derivation, we obtain the following results for the exchange correlation functions $\overline{G}^{(a_1 a_2)}_{J,E}=\overline{G}^{(a_1 a_2)}_{J,E} (j_1,j_2; \overline{\omega}_1, \overline{\omega}_2)$: 
\begin{eqnarray}
\overline{G}^{(+ +)}_{J,E} &=& -\frac{1}{N^2}\sum_{\mathbf{k}\mathbf{k}^\prime} \left\{ \left[ \frac{\mu_{\mathbf{k}}^2}{\overline{\omega}_1 - E_{\mathbf{k}}/\hbar + i\delta^+ } + \frac{|\nu_{\mathbf{k}}|^2}{\overline{\omega}_1 + E_{\mathbf{k}}/\hbar + i\delta^+ } \right] \cdot \left[ \frac{\mu_{\mathbf{k}^\prime}^2}{\overline{\omega}_2 - E_{\mathbf{k}^\prime}/\hbar + i\delta^+ } + \frac{|\nu_{\mathbf{k}^\prime}|^2}{\overline{\omega}_2 + E_{\mathbf{k}^\prime}/\hbar + i\delta^+ } \right] \right. \notag \\
&& \left. + \left[ \frac{\mu_{\mathbf{k}} \nu_{\mathbf{k}}^{*}}{\overline{\omega}_1 - E_{\mathbf{k}}/\hbar + i\delta^+ } - \frac{\mu_{\mathbf{k}} \nu_{\mathbf{k}}^{*}}{\overline{\omega}_1 + E_{\mathbf{k}}/\hbar + i\delta^+ } \right] \left[ \frac{\mu_{\mathbf{k}^\prime} \nu_{\mathbf{k}^\prime} }{\overline{\omega}_2 - E_{\mathbf{k}^\prime}/\hbar + i\delta^+ } - \frac{\mu_{\mathbf{k}^\prime} \nu_{\mathbf{k}^\prime}}{\overline{\omega}_2 + E_{\mathbf{k}^\prime}/\hbar + i\delta^+ } \right] + c.c. 
 \right\} e^{-i(\mathbf{k}-\mathbf{k}^\prime)\cdot(\mathbf{R}_{j_1}-\mathbf{R}_{j_2})} , \notag \\
&&  \label{eqna2.6} 
\end{eqnarray}
with the remaining components given by
\begin{equation}
\overline{G}^{(- -)}_{J,E} = [\overline{G}^{(+ +)}_{J,E}]^{*}, \quad  \overline{G}^{(+ -)}_{J,E} = [\overline{G}^{(- +)}_{J,E}]^{*} = 0 . \label{eqna2.7}
\end{equation}

\section{Calculation of $\overline{\langle I_1(t_1) I_2(t_2)\rangle}^{(a_1 a_2)}$ } \label{seca3}

Following a similar calculation to Appendix \ref{seca1}, we obtain the correlation functions $\overline{\langle I_1(t_1) I_2(t_2)\rangle}^{(a_1 a_2)}$ as follows. 
\begin{eqnarray}
\overline{\langle I_1(t_1) I_2(t_2)\rangle}^{(++)} &=& -\left(\frac{e T_1 T_2}{2\pi \hbar^2}\right)^2 \sum_{\sigma_1 \sigma_2 l_1 l_2} \iint d\xi_1 d\xi_2 G^{(++)}_{c,l_1 l_2}(j_1\sigma_1 t_1;  j_2\sigma_2 t_2; \frac{1}{\hbar}(\xi_1 - eV_1), \frac{1}{\hbar}(\xi_2 - eV_2)) G^{(l_1)}_{d_1}(\xi_1) G^{(l_2)}_{d_2}(\xi_2) \notag \\
&& \times A_{d_1}(\xi_1) A_{d_2}(\xi_2) , \label{eqna3.1}
\end{eqnarray}
where 
\begin{equation}
G^{(++)}_{c,l_1 l_2}(j_1\sigma_1 t_1; j_2\sigma_2 t_2; \omega_1, \omega_2) = \iint_c d t_2^\prime d t_1^\prime \overline{\theta}^{(l_1)}_c(t_1^\prime-t_1) \overline{\theta}^{(l_2)}_c(t_2^\prime-t_2) G_c(j_1\sigma_1 t_1, j_2\sigma_2 t_2; j_2\sigma_2 t_2^\prime, j_1\sigma_1 t_1^\prime) e^{-i\omega_1(t_1^\prime-t_1)-i\omega_2 (t_2^\prime-t_2)} . \label{eqna3.2} 
\end{equation}
\begin{eqnarray}
\overline{\langle I_1(t_1) I_2(t_2)\rangle}^{(--)} &=& -\left(\frac{e T_1 T_2}{2\pi \hbar^2}\right)^2 \sum_{\sigma_1 \sigma_2 l_1 l_2} \iint d\xi_1 d\xi_2 G^{(--)}_{c,l_1 l_2}(j_1\sigma_1 t_1;  j_2\sigma_2 t_2; \frac{1}{\hbar}(\xi_1 - eV_1), \frac{1}{\hbar}(\xi_2 - eV_2)) G^{(l_1)}_{d_1}(\xi_1) G^{(l_2)}_{d_2}(\xi_2) \notag \\
&& \times A_{d_1}(\xi_1) A_{d_2}(\xi_2) , \label{eqna3.3}
\end{eqnarray}
where 
\begin{equation}
G^{(--)}_{c,l_1 l_2}(j_1\sigma_1 t_1; j_2\sigma_2 t_2; \omega_1, \omega_2) = \iint_c d t_2^\prime d t_1^\prime \overline{\theta}^{(l_1)}_c(t_1-t_1^\prime) \overline{\theta}^{(l_2)}_c(t_2-t_2^\prime) G_c(j_1\sigma_1 t_1^\prime, j_2\sigma_2 t_2^\prime; j_2\sigma_2 t_2, j_1\sigma_1 t_1) e^{i\omega_1(t_1^\prime-t_1)+i\omega_2 (t_2^\prime-t_2)} . \label{eqna3.4} 
\end{equation}
\begin{eqnarray}
\overline{\langle I_1(t_1) I_2(t_2)\rangle}^{(+-)} &=& \left(\frac{e T_1 T_2}{2\pi \hbar^2}\right)^2 \sum_{\sigma_1 \sigma_2 l_1 l_2} \iint d\xi_1 d\xi_2 G^{(+-)}_{c,l_1 l_2}(j_1\sigma_1 t_1;  j_2\sigma_2 t_2; \frac{1}{\hbar}(\xi_1 - eV_1), \frac{1}{\hbar}(\xi_2 - eV_2)) G^{(l_1)}_{d_1}(\xi_1) G^{(l_2)}_{d_2}(\xi_2) \notag \\
&& \times A_{d_1}(\xi_1) A_{d_2}(\xi_2) , \label{eqna3.5}
\end{eqnarray}
where 
\begin{equation}
G^{(+-)}_{c,l_1 l_2}(j_1\sigma_1 t_1; j_2\sigma_2 t_2; \omega_1, \omega_2) = \iint_c d t_2^\prime d t_1^\prime \overline{\theta}^{(l_1)}_c(t_1^\prime-t_1) \overline{\theta}^{(l_2)}_c(t_2-t_2^\prime) G_c(j_1\sigma_1 t_1, j_2\sigma_2 t_2^\prime; j_2\sigma_2 t_2, j_1\sigma_1 t_1^\prime) e^{-i\omega_1(t_1^\prime-t_1)+i\omega_2 (t_2^\prime-t_2)} . \label{eqna3.6} 
\end{equation}
\begin{eqnarray}
\overline{\langle I_1(t_1) I_2(t_2)\rangle}^{(-+)} &=& \left(\frac{e T_1 T_2}{2\pi \hbar^2}\right)^2 \sum_{\sigma_1 \sigma_2 l_1 l_2} \iint d\xi_1 d\xi_2 G^{(-+)}_{c,l_1 l_2}(j_1\sigma_1 t_1;  j_2\sigma_2 t_2; \frac{1}{\hbar}(\xi_1 - eV_1), \frac{1}{\hbar}(\xi_2 - eV_2)) G^{(l_1)}_{d_1}(\xi_1) G^{(l_2)}_{d_2}(\xi_2) \notag \\
&& \times A_{d_1}(\xi_1) A_{d_2}(\xi_2) , \label{eqna3.7}
\end{eqnarray}
where 
\begin{equation}
G^{(-+)}_{c,l_1 l_2}(j_1\sigma_1 t_1; j_2\sigma_2 t_2; \omega_1, \omega_2) = \iint_c d t_2^\prime d t_1^\prime \overline{\theta}^{(l_1)}_c(t_1-t_1^\prime) \overline{\theta}^{(l_2)}_c(t_2^\prime-t_2) G_c(j_1\sigma_1 t_1^\prime, j_2\sigma_2 t_2; j_2\sigma_2 t_2^\prime, j_1\sigma_1 t_1) e^{i\omega_1(t_1^\prime-t_1)-i\omega_2 (t_2^\prime-t_2)} . \label{eqna3.8} 
\end{equation}

\section{Calculation of $G^{(2)}_{J,E}$ in Fermi liquid state for extended double-tip STS } \label{seca4}

We now evaluate the exchange contribution $G^{(2)}_{J,E}$ in the Fermi liquid state for the extended double-tip STS. Following an analogous decomposition to Eq. (\ref{eqn3.2.7}), we express this quantity as a sum of four components 
\begin{equation}
G^{(2)}_{J,E}(j_1 \sigma_1 t_1, j_2 \sigma_2 t_2; \omega_1, \omega_2) = \sum_{a_1 a_2 =\pm} G^{(a_1 a_2)}_{J,E} (j_1 \sigma_1 t_1, j_2 \sigma_2 t_2; \omega_1, \omega_2), \label{eqna4.1}
\end{equation}
where each component is given by the contour-ordered correlation function
\begin{equation}
G^{(a_1 a_2)}_{J,E}(j_1\sigma_1 t_1,j_2\sigma_2 t_2; \omega_1, \omega_2) = \iint_c d t_2^\prime d t_1^\prime
\wick{ \langle T_c J^{(a_1)}_{1A} (j_1\sigma_1; \c1 t_1, \c2 t_1^\prime; \omega_1) J^{(a_2)}_{2A} (j_2\sigma_2; \c2 t_2, \c1 t_2^\prime; \omega_2) \rangle }. \label{eqna4.2}
\end{equation}

Applying the methodology developed in Sec. \ref{sec3.2}, we obtain the explicit forms of these components as
\begin{eqnarray}
&& G^{(++)}_{J,E}(j_1\sigma_1 t_1,j_2\sigma_2 t_2; \omega_1, \omega_2) = -\delta_{\sigma_1 \sigma_2} e^{i \Delta	\omega \cdot \Delta t} \frac{1}{N^2} \sum	_{\mathbf{k}\mathbf{k}^\prime} \frac{e^{i(\mathbf{k}-\mathbf{k}^\prime )\cdot(\mathbf{R}_{j_1}-\mathbf{R}_{j_2} )}}{(\omega_1 - \varepsilon_{\mathbf{k}^\prime}/\hbar + i \delta^+)(\omega_2 - \varepsilon_{\mathbf{k}}/\hbar + i\delta^+)}  , \label{eqna4.3-1} \\
&& G^{(--)}_{J,E}(j_1\sigma_1 t_1,j_2\sigma_2 t_2; \omega_1, \omega_2) = -\delta_{\sigma_1 \sigma_2} e^{-i \Delta	\omega \cdot \Delta t} \frac{1}{N^2} \sum	_{\mathbf{k}\mathbf{k}^\prime} \frac{e^{i(\mathbf{k}-\mathbf{k}^\prime )\cdot(\mathbf{R}_{j_1}-\mathbf{R}_{j_2} )}}{(\omega_1 - \varepsilon_{\mathbf{k}}/\hbar - i\delta^+)(\omega_2 - \varepsilon_{\mathbf{k}^\prime}/\hbar - i \delta^+)}  , \label{eqna4.3-2} \\
\end{eqnarray}
while the remaining components vanish 
\begin{equation}
G^{(+-)}_{J,E}(j_1\sigma_1 t_1,j_2\sigma_2 t_2; \omega_1, \omega_2) = 0, \quad G^{(-+)}_{J,E}(j_1\sigma_1 t_1,j_2\sigma_2 t_2; \omega_1, \omega_2)=0 . \label{eqna4.4}
\end{equation}
Here we have defined $\Delta \omega = \omega_2 - \omega_1$ and $\Delta t = t_2 - t_1$. 

\end{widetext}



%

\end{document}